\documentclass{ieeeaccess}
\usepackage[T1]{fontenc}
\usepackage{cite}
\usepackage{amsmath,amssymb,amsfonts}
\usepackage{algorithmic}
\usepackage{graphicx}
\usepackage{textcomp}
\usepackage[normalem]{ulem}

\usepackage{bm}
\usepackage{url}
\makeatletter
\AtBeginDocument{\DeclareMathVersion{bold}
\SetSymbolFont{operators}{bold}{T1}{times}{b}{n}
\SetSymbolFont{NewLetters}{bold}{T1}{times}{b}{it}
\SetMathAlphabet{\mathrm}{bold}{T1}{times}{b}{n}
\SetMathAlphabet{\mathit}{bold}{T1}{times}{b}{it}
\SetMathAlphabet{\mathbf}{bold}{T1}{times}{b}{n}
\SetMathAlphabet{\mathtt}{bold}{OT1}{pcr}{b}{n}
\SetSymbolFont{symbols}{bold}{OMS}{cmsy}{b}{n}
\renewcommand\boldmath{\@nomath\boldmath\mathversion{bold}}}
\makeatother

\def\BibTeX{{\rm B\kern-.05em{\sc i\kern-.025em b}\kern-.08em
    T\kern-.1667em\lower.7ex\hbox{E}\kern-.125emX}}

\usepackage{subfig}
\begin{document}
\history{This work has been accepted for publication in IEEE Access. The final published version will be available via IEEE Xplore: \url{https://doi.org/10.1109/ACCESS.2026.3696825}}
\doi{https://doi.org/10.1109/ACCESS.2026.3696825}

\title{Integrating Network and Attack Graphs for Service-Centric Impact Analysis}

\author{Joni Herttuainen\authorrefmark{1},
Vesa Kuikka\authorrefmark{1},
and Kimmo K. Kaski\authorrefmark{1}
}

\address[1]{Department of Computer Science, Aalto University School of Science,
P.O. Box 11000, 00076 Aalto, Finland (e-mail: firstname.lastname@aalto.fi)}

\tfootnote{This work has been accepted for publication in IEEE Access. The final published version will be available via IEEE Xplore: \url{https://doi.org/10.1109/ACCESS.2026.3696825}}

\markboth
{Herttuainen \headeretal: Integrating Network and Attack Graphs for Service-Centric Impact Analysis}
{Herttuainen \headeretal: Integrating Network and Attack Graphs for Service-Centric Impact Analysis}

\corresp{Corresponding author: Joni Herttuainen (e-mail: joni.herttuainen@aalto.fi)}

\begin{abstract}
Cyberattacks on enterprise networks exploit complex dependencies among infrastructure, services, and applications, which challenge traditional analysis methods that focus on attack paths or network topology in isolation. In this study, we introduce a novel probabilistic multilayer modelling framework, based on influence propagation in networks, that integrates attack graphs with the communication network topology, enabling a service-centric impact analysis of cyberattacks. Our method captures both the vulnerability exploitability and network connectivity, allowing us to assess the likelihood of attack propagation and cumulative impacts across interconnected services. By integrating standard vulnerability metrics (such as CVSS) with the network-level connectivity probabilities, the framework provides a cohesive view of the dynamics of cyberattacks.
We validate this approach using a realistic case study of an enterprise network,
demonstrating its ability to determine critical nodes, vulnerabilities, and service dependencies that significantly influence attack outcomes. Our findings show that integrating network and attack graph perspectives offers more actionable insights into
risk assessment and mitigation planning, advancing the analysis of cyberattacks in complex networked environments.
\end{abstract}


\titlepgskip=-21pt

\maketitle

\section{Introduction}
\label{sec:Introduction}

\PARstart{U}{nderstanding} the nature of cyber threats and protecting systems from attacks has become increasingly critical.
Cyberattacks occur when malicious actors compromise the confidentiality, integrity, or availability of computer networks and/or services, by exploiting vulnerabilities therein~\cite{li2021comprehensive,wachter2023graph}.
Cyberattackers typically gain access to computer networks by using a series of exploits, in which each exploit facilitates the next one. Once they have an initial entry point, they proceed with lateral movement to spread to the rest of the network and reach other hosts and applications within an organization~\cite{SMILIOTOPOULOS2024e26317}.
This series of exploits constitutes an attack path, while the collection of potential attack paths is an attack graph~\cite{zenitani2023attack, lallie2020review, zeng2019survey}.

In order to reduce the consequences of these attacks when they occur, there is a need for mitigation strategies to address the vulnerabilities of the computer network.
Although perfect security is unattainable, the implementation of defense mechanisms can make systems less susceptible to cyber-targeting.
In this context, modelling cyber threats and various attack scenarios can serve as a valuable tool for analysing known vulnerabilities and evaluating their impact on the system, thereby supporting the deployment of protective strategies.

To enhance network security planning, it is crucial to first evaluate system vulnerabilities and understand how attacks can spread through interconnected components. Model-based approaches have been developed to facilitate this analysis. For instance, the framework outlined in \cite{lof2010approach, CySeMoL} utilizes probabilistic modelling to represent attack paths and countermeasures, allowing for the estimation of
success probabilities of an attack across various architectural scenarios. Likewise, the research presented in \cite{lippmann2006validating} explores strategies for safeguarding critical resources within enterprise networks by employing layered firewall architectures, which include both perimeter and internal firewalls to achieve subnet isolation.

While these methods offer valuable insights into network defense and the likelihood of attacks, they primarily concentrate on either probabilistic attack modelling or network segmentation. They do not effectively capture how attacks propagate across services or quantify their cumulative effects on user-facing functionalities. This highlights the need for an integrated approach that combines attack propagation, network structure, and impact on service levels.

In the present study, we investigate cyberattacks using a novel approach that combines network modelling and graphs~\cite{noel2018review,lagraa2024review,idika2010extending}.
This results in an attack graph that is a visual representation of potential paths an attacker could take to compromise a system~\cite{ou2011quantitative,stergiopoulos2022automatic,gain2023attack}.
Attack graphs are commonly represented as directed acyclic graphs (DAGs), where nodes and links describe the structure of possible attack paths~\cite{kordy2014dag}. Each node in the attack graph can represent a host, network device, or vulnerability of a network system~\cite{zeng2019survey,Ou2011}.
In this work, we adopt an acyclic attack graph representation, which ensures tractable computation of attack propagation probabilities. However, we note that the underlying influence propagation framework used in this study is more general and has previously been shown to support cyclic dependencies~\cite{kuikkaSciRep, kuikka2022efficiency}. While such extensions are not required for the present analysis, this property enables future applications to more complex scenarios where cyclic interactions between system components may arise.

There are two commonly used types of attack graphs. The first one is a directed graph in which nodes denote system states related to security, such as file modifications, levels of user privileges, and trust relationships among hosts.
The edges represent possible exploits or actions by the perpetrators.
The second type involves nodes that denote the preconditions or postconditions of an exploit, with edges showing the effects of those conditions.
An attack surface of a system refers to the various ways an intruder can exploit its weaknesses to gain access,~\cite{ou2011quantitative} including the resources that a perpetrator can access.
The more resources an attacker can access, the larger the attack surface is, making the system less secure.
To evaluate the security of a system based on its attack surface, several metrics have been developed~\cite{AttackSurfaceMetric}.

In the present study, we introduce a quantitative attack modelling approach to combine vulnerability metrics of an enterprise network by integrating the attack graph and the structure of the communication network. In our probabilistic approach, we use exploitability metrics to evaluate the likelihood that individual service vulnerabilities will be exploited along the attack paths~\cite{lof2010approach,wang2008attack}.
The model estimates the overall probability that a perpetrator successfully gains specific privileges or executes attacks.
These probabilities are determined by merging alternative joining attack paths using probability theory for non-exclusive events~\cite{kuikkaSciRep}. By combining probabilities with vulnerability impact scores, our approach estimates the expected cumulative impact across various states of the attack and at the service level.

This approach is novel, as it employs a detailed-level network modelling technique~\cite{kuikkaSciRep} and integrates the communication network and attack graph structures. It enables the construction of a quantitative model that describes cyber threats and attacks in the communication network. The outcomes of the model can be presented at varying levels of detail, incorporating attack graph nodes that represent microservices, communication network nodes like servers and network equipment, as well as user services and applications.
Our approach is based on multilayer network models~\cite{bianconi2018multilayer,interdonato2020multilayer} and allows us to analyse cyber threats and design protective solutions against cyberattacks in enterprise networks. Here, we focus on investigating the infrastructure and communication layer, the cyberattack layer, and the application layer, each of which can be further divided into sublayers. With the present approach, we can also evaluate the operation and robustness of the system from the point of view of physical network, protocol layers, cyber incidents, user services, and applications, all of which have the potential to simultaneously exert detrimental impacts on system users.

Moreover, we construct an additional abstraction layer for each user-facing service, to enable the evaluation of metrics related to incoming and outgoing attack effects.
In this context, a user-facing service refers to an application or functionality provided to end users (e.g., web services, databases, or microservices), rather than individual human users. These service-level layers group together the underlying network nodes and vulnerabilities that support a given service, allowing us to analyse how attacks propagate within and between services. This abstraction enables the quantification of both the impact of incoming attacks on a service and the potential for that service to propagate attacks to other parts of the system.
This in turn allows us to analyse cyber threats targeting specific services or applications as well as the way an attack propagates within a service between its microservices. Different services or application servers provide different options for detection, which affects the likelihood of successful mitigation. It is an essential feature of our method that individual attack paths and attack graphs can be extended to different services and servers in the network topology.

In this study, our overarching aim is to create a comprehensive methodology to model how cyberattacks affect user services within an enterprise network architecture.
By integrating probabilistic methods to trace attack propagation in attack graphs with user service and application tiers in communication networks, the results of the model help us analyse cyber threats and attacks. Furthermore, microservice vulnerabilities can be linked to user-level service functionalities, which in turn facilitates the design of optimal defense and mitigation strategies. Our modelling approach allows us to explore different cyberattack scenarios and relevant protection strategies against them.

This study is organised as follows. In Section~\ref{sec:RelatedWork} we discuss related works. Next, in Section~\ref{sec:Method} we introduce our network-based modelling methodology and related metrics to analyse the impact of cyberattacks. In Section~\ref{sec:Case_Study} we present the results of the impact analysis of cyberattacks against services in a realistic communication network under different scenarios to find ways to mitigate them. The final conclusions are drawn in Section~\ref{sec:Conclusion}.

\section{Related Work}
\label{sec:RelatedWork}

In this section, we present a brief overview of related research on modelling cyberattacks and analysing their impact, risk, and resilience in enterprise networks.
Here, we focus specifically on graph-based models, attack propagation models, and multilayer network models as well as the metrics to analyse network security~\cite{wang2017network,homer2013aggregating,pendleton2016survey}.
In~\cite{homer2013aggregating} the authors present a probabilistic metric for network security by aggregating individual vulnerability measures derived from the Attack Complexity sub-metric of the Common Vulnerability Scoring System~\cite{Scoring} (CVSS) in total network score, whereas in~\cite{wang2008attack}, the overall likelihood of compromise is calculated using probabilities derived from the CVSS Base Score.
The survey on system security metrics presented in~\cite{pendleton2016survey} proposes a framework to measure system-level security by using the following four sub-metrics: system vulnerabilities, defense capabilities, attack severity, and situational factors.
Here the authors introduce a hierarchical ontology to evaluate relationships among these sub-metrics.
They also review existing security metrics and highlight their advantages and limitations to guide the development of more effective security metrics.
The study in~\cite{Noel2017} introduces a set of normalised metrics for comparisons across enterprises.
It categorises related metrics into four families (victimization, network size, containment, and topology) that aggregate individual scores into family scores.
These metrics are presented through interactive visualisations that show trends over time at individual, family, and overall network levels.

In order to analyse a particular part of the network resilience, we need to apply network modelling techniques and graph models to characterise cyberattacks~\cite{noel2018review,lagraa2024review}.
The probability formulas that are used to describe connectivity are also useful in explaining the concept of resilience~\cite{Ball_etal, kuikka2022resilience}.
Here, the connectivity between two nodes is defined as the probability that there is a functioning connection between them. If the connectivity values between all pairs of neighbouring nodes in the network are known, the connectivity probabilities between any pair of nodes in the network can be calculated~\cite{Ball_etal}. This calculation can be performed using either simulation methods or algorithmic techniques.
An alternative method to this was introduced in~\cite{Almiala_Aalto_Kuikka}, where probabilities were calculated based on spreading probabilities through self-avoiding paths with no repeated nodes.

In the graph-based model of~\cite{stergiopoulos2022automatic}, the edge weights indicate potential risks.
As a result, an edge with the highest weight represents the worst-case scenario.
Depending on the specific problem, the optimal edge or graph can be determined based on the maximum or minimum weight~\cite{stergiopoulos2022automatic}.
Another research~\cite{ammann2005host} introduced a host-host model to assess network vulnerabilities and the attack chaining process.
This model adopts a penetration testing perspective to determine the maximum level of compromise achievable on a host.

In~\cite{noel2009proactive} authors consider proactive intrusion prevention and response using attack graphs, dividing them into unconstrained or constrained graphs.
Imposing constraints leads to a reduced scope, thus simplifying complexity and improving understanding of potential threats. An unconstrained graph includes all conceivable attack paths without predefined origins or destinations, presuming that the threat source is unidentified and no targeted assets are specified. In contrast, in a constrained graph, particular starting points are defined, including only those paths that are accessible from the origins. The graph can also end with specified attack goals, concentrating on the protection of critical network assets or the attack surfaces of services or applications employed by users. Several attack surface metrics~\cite{ou2011quantitative,stuckman2012comparing,ayrour2018modelling} have been developed to assess an attacker's capability to compromise a system or its assets.

The enterprise networks are multilayer systems that represent interconnected entities of various types, i.e. networks of nodes, edges, and layers of different types of interactions~\cite{interdonato2020multilayer,hammoud2020multilayer,hong2016towards}.
To facilitate analysis, multiple methods have been proposed to simplify and contract these networks.
The study in~\cite{interdonato2020multilayer} presents a comprehensive taxonomy of existing simplification techniques along with the classification of multilayer network methods.

While prior work has studied attack graphs and network resilience, they are often considered in isolation, and their interrelationship has received limited attention.
Network resilience focuses on the ability of a network to maintain connectivity under failures or disruptions, whereas attack graphs model how vulnerabilities can be exploited to propagate attacks.
These perspectives are inherently linked, as attack propagation depends not only on vulnerabilities but also on the availability of communication paths between system components.
However, existing approaches, such as the probabilistic methods of~\cite{wang2008attack, homer2013aggregating}, the attack graph metric suite of~\cite{Noel2017}, and the method of~\cite{stergiopoulos2022automatic} based on closeness centrality and dependency paths, do not consider per-service impact metrics or model the communication network alongside the attack graph.
We propose a service-centric, integrated multilayer modelling approach to address these limitations.

\section{Method}
\label{sec:Method}

Our methodology calculates attack propagation probabilities and service-centric impact metrics from a graph obtained by combining a network graph and an attack graph.
While a network graph, representing the internode communication in the network, can be derived directly from the network topology either manually or with the help of automated tools, the attack graph creation is a more complex process.
Although there are automated tools that can be utilized to identify software vulnerabilities in the network, understanding how they can be exploited and how a cyberattack propagates through the network requires expertise.
Therefore, constructing an attack graph is often a manual task carried out by cybersecurity experts.

In our model, an attack graph consists of nodes representing the attack states reached by exploiting the vulnerabilities of the system.
The vulnerability is denoted by a directed edge between two attack states.
Each edge is assigned a weight $p_E \in [0, 1]$ that denotes the likelihood of exploiting the vulnerability in an attack.
In the present study, the values are derived from the CVSS exploitability scores following \cite{kuikka2024networkmodelling}, but any suitable exploitability estimate can be used. In addition, the attack graph has certain start and end states, which roughly correspond to an \emph{attacker being ready to attack} and \emph{attacker having reached the desired goal(s)}, respectively.

\begin{figure}[htpb]
    \begin{minipage}{\columnwidth}
        \centering
        \begin{minipage}{.45\textwidth}
            \subfloat[]{
                \includegraphics[width=\textwidth,trim=5cm 0.5cm 5cm 0.5cm,clip]{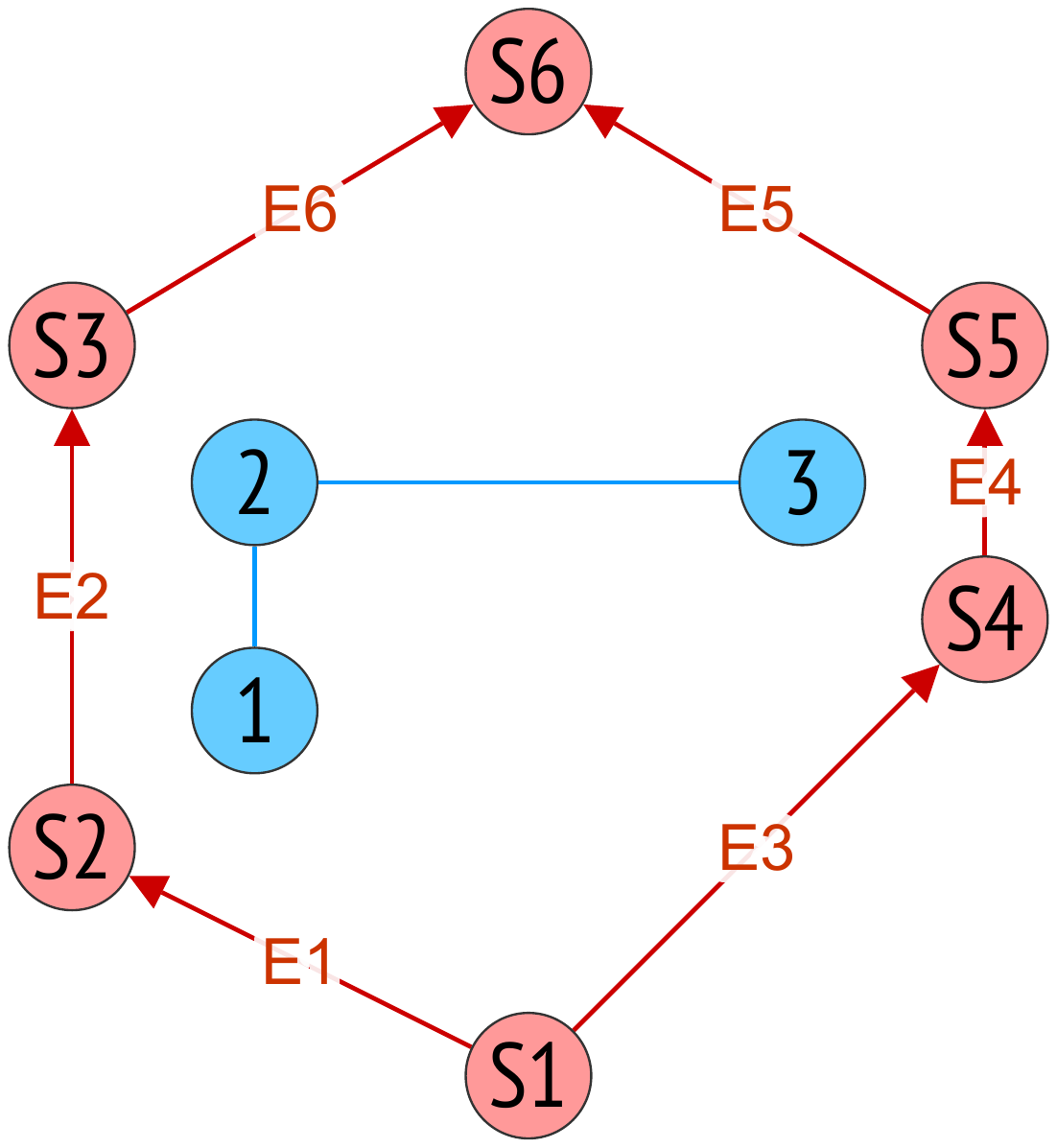}
                \label{sfig:example_init}
            }
        \end{minipage}
        \begin{minipage}{.45\textwidth}
            \subfloat[]{
                \includegraphics[width=\textwidth,trim=5cm 0.5cm 5cm 0.5cm,clip]{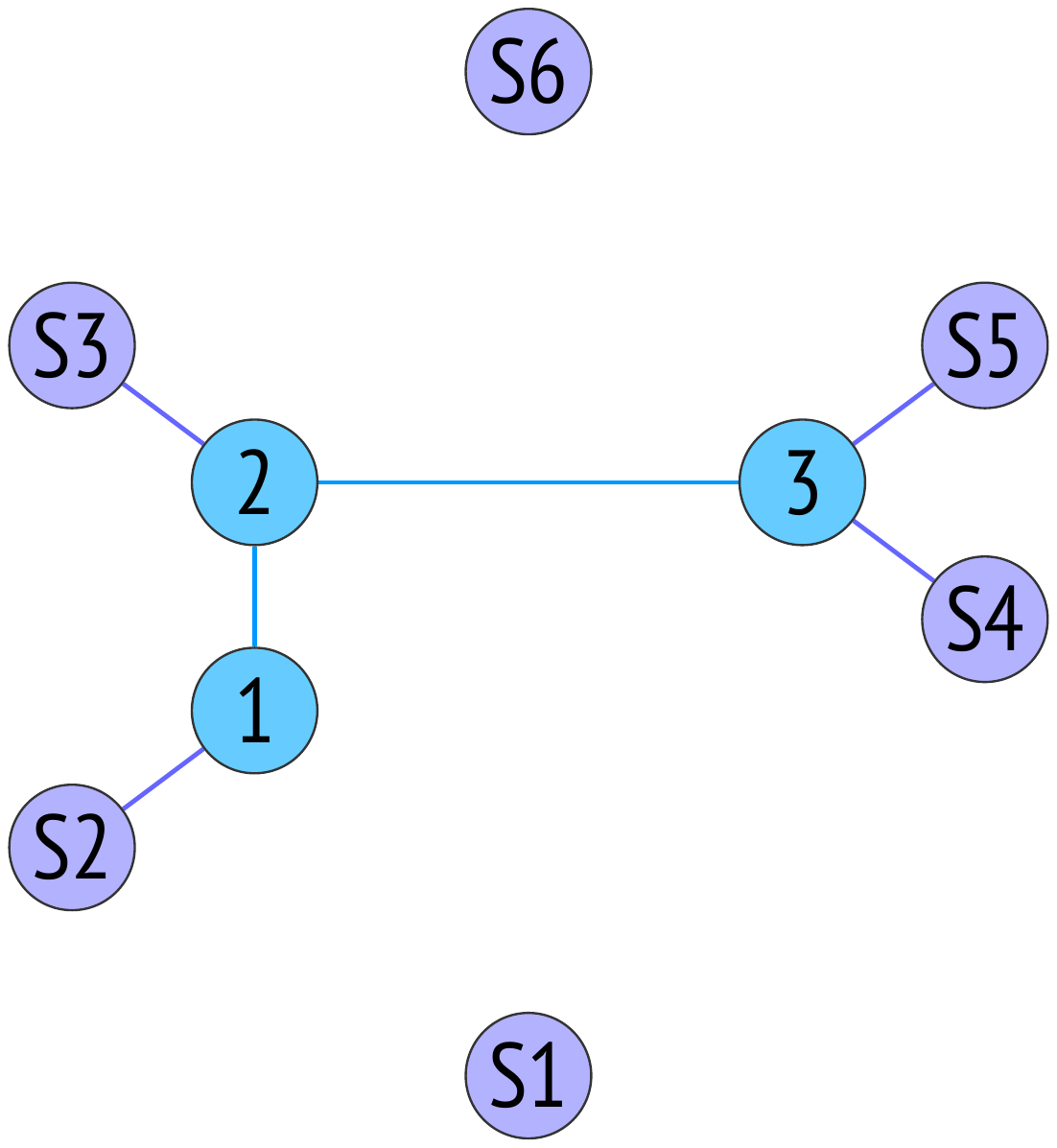}
                \label{sfig:example_states_attached}
            }
        \end{minipage}
        \begin{minipage}{.45\textwidth}
            \subfloat[]{
                \includegraphics[width=\textwidth,trim=5cm 0.5cm 5cm 0.5cm,clip]{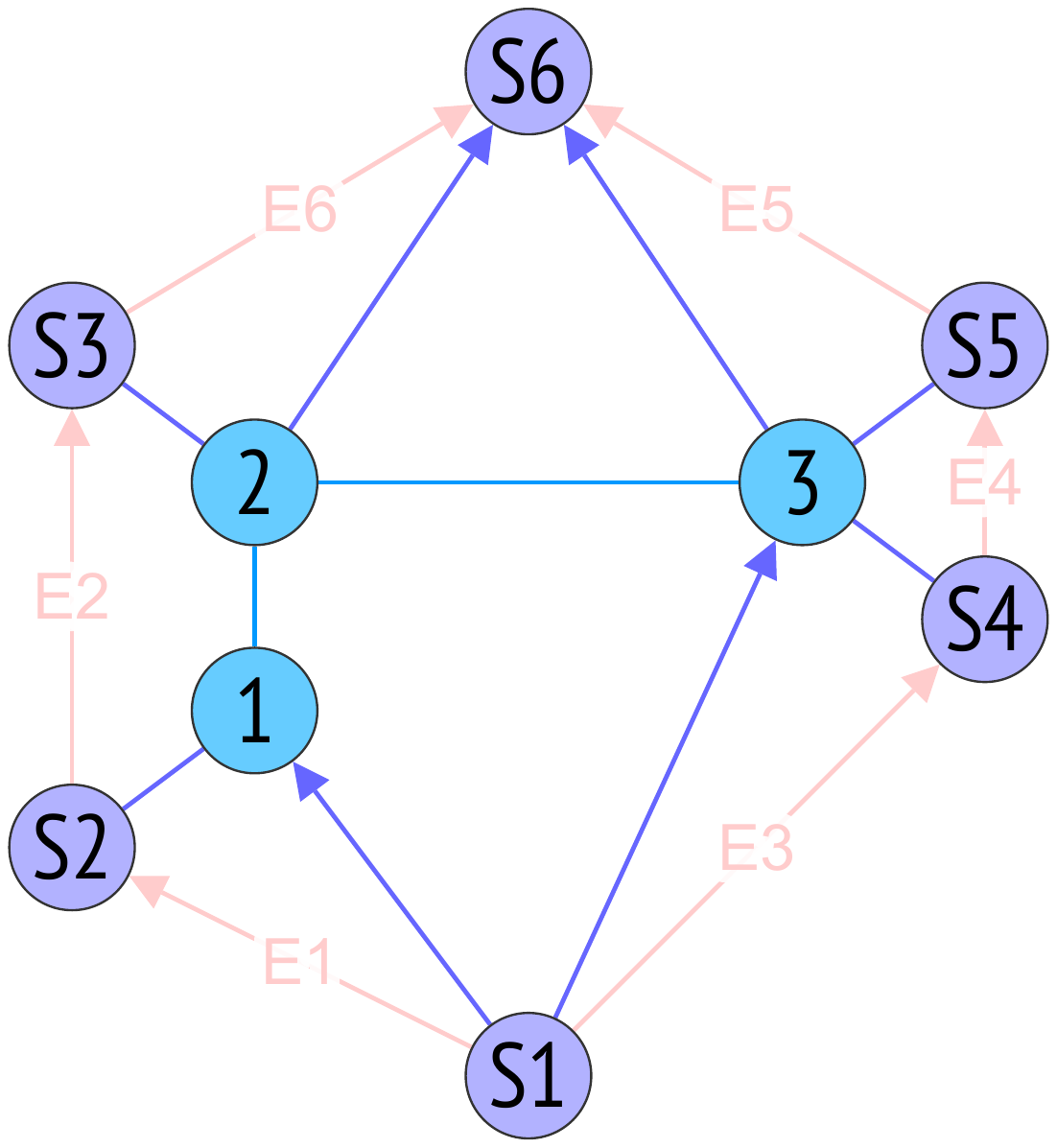}
                \label{sfig:example_combined}
            }
        \end{minipage}
        \caption[]{
            Combining network graph and attack graph.
            (Here the graphs are only for demonstrative purposes, not the ones used in the rest of the study.)
            \subref{sfig:example_init} Original network and attack graph in blue and red colour, respectively.
            \subref{sfig:example_states_attached} Attack states connected to their affected services.
            \subref{sfig:example_combined} Combined graph with connected start and end states $S1$ and $S6$, respectively.
        }
        \label{fig:example_combine_net_attack}
    \end{minipage}
\end{figure}

The attack graph is combined with the network graph of the underlying communication network by grouping the attack states by the affected services represented by nodes in the network graph, as depicted in Fig.~\ref{fig:example_combine_net_attack}.
As the start and end states do not represent the state of a particular service, they are left out of the grouping.
The grouped attack states are then added to the network graph and connected to the corresponding nodes --- on which the services are running~--- with \emph{undirected} edges, as presented in Fig.~\ref{sfig:example_states_attached}.

The start and end states are connected to the relevant network nodes with \emph{directed} links following the flow of an attack as shown in Fig.~\ref{sfig:example_combined}.
The start state is connected to those network nodes whose attack states are directly reachable from the start state, and the end state is connected from those network nodes whose attack states can directly reach the end state in the attack graph.
All added directed and undirected links are set to the constant weight of $1.0$, which denotes the probability of 100\% connection between the attack states and the nodes to which they connect. The vulnerabilities ($EN$), as represented in the attack graph, are not actually present in the combined graph but are left in Fig.~\ref{sfig:example_combined} with decreased opacity for demonstration purposes.

The link weights $w_n \in [0, 1]$ of a network graph represent the probability of a successful communication between connected nodes. They can also be interpreted as, e.g., the probabilities of working routing between the nodes, the probabilities of the attack passing the network intrusion detector or a combination of different probabilities. Since the practical estimation of $w_n$ is outside the scope of this study, we vary its value to explore its influence on the model.

For the purposes of demonstrating behaviour of our model, we assume that the network-level probabilities are aggregated into single values and, as is common in physical network infrastructures that the pairwise node-to-node probabilities are equal in both directions, although neither is a restriction of the model.

Our analysis of an attack is based on both the attack graph and the combined graph.
While the attack graph can be used for the analysis of exploitability and impact of an attack, the combined graph takes into account the possibility of the attack at the network level.
For example, in Fig.~\ref{sfig:example_init} there are two possible attack paths: $(E1,E2,E6)$ and $(E3,E4,E5)$.
If for any reason the service in node~2 is unreachable from node~1, the former attack path is no longer possible.
This can be clearly seen in the combined graph in Fig.~\ref{sfig:example_combined}: the vulnerability $E2$ cannot be exploited if node~2 and, thus, the attack state $S3$ is unreachable from $S2$.
However, the other attack path traversing through the service in node~3 is still possible as it is independent of the service in node~2.

Using the combined graph, we calculate the probabilities of each vulnerability $EN$ present in the attack graph at the network level.
That is, for each attack step
$Si{\rightarrow}Sj$, the corresponding network-level probability ($p_N$) of reaching the target attack state from the source attack state via any non-intersecting path through the network is calculated using the Simple Contagion Algorithm presented in \cite{kuikka2022efficiency}.
The combined probability of a successful exploitation of a vulnerability between two states is then calculated as the product of the probability that the attack step passes the underlying network and the probability of the vulnerability being exploited:
\begin{equation} \label{eq:prob_ovr}
    p(i,j) = p_{N}(i,j) * p_{E}(i,j).
\end{equation}
Here, $p_N$ and $p_E$ are the probability of an attack from state $i$ to state $j$ passing through the network and the exploitability of the vulnerability connecting the states (as depicted in the attack graph), respectively.
Since the attack states are connected to the network nodes with links of weight 1.0, regardless of the network link weights $w_n$, the network-level probability of a connection between the states is equal to the probability of a connection between their corresponding nodes.
That is:
\begin{equation*}
    \forall i \in G_m,\ j \in G_n: p_N(i,j) = p_N(m,n),
\end{equation*}
where $i$ and $j$ are the source and target attack states and $G_m$ and $G_n$ are sets consisting of attack states grouped to services running on nodes $m$ and $n$ in the network, respectively. Moreover, since the attack can traverse any non-intersecting path through the network, if $w_n<1$, the more possible paths there are between the states, the larger the $p_N$ is.

In order to measure the importance of different
vulnerabilities and services, we use metrics based on the probability, expected impact, and expected cumulative impact of an attack. Our interpretation and calculations of the probabilities of different states in the attack graph are based on the principles presented in~\cite{wang2008attack}.
First of all, our model assumes that the probability of a vulnerability being exploited (i.e., exploitability) represents the probability that a vulnerability will be exploited during an attack.
Secondly, we assume that a vulnerability can only be exploited if the attack has already propagated to the preceding state.
Thirdly, our model considers the different paths leading to a state independent, which means that the probability of an attack reaching a state is equal to the probability that at least one of the attack paths leading to the state is traversed during an attack.

The overall probability of a single vulnerability being successfully exploited during an attack is calculated as:
\begin{equation*}
    P(i,j) = P(i) * p(i,j),
\end{equation*}
where $p(i,j)$ is the probability of a vulnerability between states $i$ and $j$ being successfully exploited (see:~\eqref{eq:prob_ovr}) and $P(i)$ is the probability that an attack has reached the state $i$.
Since the different paths leading to a state are considered independent, $P(i)$ is calculated as:
\begin{equation*}
    P(i) = 1 - \prod_{j \in A} [1 - P(j,i)],
\end{equation*}
where $A$ is the set of all states immediately preceding state $i$.

Our calculation of expected impact is based on the expectation from probability theory:
\begin{equation*}
    ExpectedImpact = Probability * Impact,
\end{equation*}
which is motivated by the calculation of risk as a function of likelihood and severity in~\cite{stergiopoulos2022automatic} and assessment of risk as a combination of likelihood and impact in~\cite{nist2012article}.
The impact is a property of the vulnerabilities between the attack states and not of the attack states themselves.
Furthermore, the impact values of the different vulnerabilities leading to a state may differ from each other.
To account for this, we consider the expected impact $E[I_{i}]$ of reaching a state as a sum of the expected impacts of the vulnerabilities $E[I_{j,i}]$ leading to the state~$i$:
\begin{equation} \label{eq:E_IMP}
    E[I_{i}] = \sum_{j \in A} E[I_{j,i}] = \sum_{j \in A} P(j,i)*I(j,i),
\end{equation}
where $I(j,i)$ is the impact of a vulnerability from state $j$ to state $i$ and $A$ is the set of all states $j$ immediately preceding the state $i$.
Now we can calculate the expected cumulative impact of an attack reaching a certain state:

\begin{equation} \label{eq:CEI}
    CE[I_{i}] = E[I_{i}]+ \sum_{j \in B} E[I_{j}],
\end{equation}
where $B$ is the set of all the states on every possible path from the beginning of an attack to state $i$, excluding the state $i$ itself.

Next we calculate three other metrics that are motivated by the grouping of attack states in their affected services, namely expected cumulative inbound and outbound impacts, as well as expected node-wise impact. The expected cumulative inbound impact represents the case in which the attack has propagated to a certain service and gained a foothold there but has not yet propagated further within the service.
It is calculated as a sum of the expected impacts of all the preceding states and the expected impacts (see:~\eqref{eq:E_IMP}) of the incoming vulnerabilities to the service:
\begin{equation} \label{eq:ICEI}
    C_{IN}(n) = \sum_{i \in C} (E[I_{i}] + \sum_{j \in G_n} E[I_{i,j}]),
\end{equation}
where $C$ is the set of all states on every possible path from the beginning of an attack to the states in $G_n$, excluding the states in $G_n$,
which is the set of attack states grouped by the service running on node $n$. The expected cumulative outbound impact can be calculated
by extending $C_{IN}$:
\begin{equation} \label{eq:OCEI}
    C_{OUT}(n) = C_{IN}(n) + \sum_{\substack{i \in G_n \\ j \in D}} E[I_{i,j}],
\end{equation}
where $D$ is the set of all states immediately following any state in $G_n$, including those in $G_n$, which is the set of attack states grouped by the service $n$.
This represents the case in which a service is used to gain a foothold in another service or to fulfill the criteria set for a successful attack (i.e., to reach the end state).

The node-wise impact represents the maximum impact that can be caused to
a single service if all the vulnerabilities of the service are exploited.
It is calculated as a sum of the expected impacts of vulnerabilities affecting a particular node without taking into account the impact cumulated prior to the attack reaching the service:
\begin{equation} \label{eq:NWCEI}
    C_{NODE}(n) =  \sum_{i \in G_n} E[I_{i}],
\end{equation}
where $G_n$ is the set of attack states grouped in service~$n$.

An illustrative example of the vulnerabilities and states included in each of the metrics is presented in Appendix~\ref{app:metrics}.

\section{Case Study}
\label{sec:Case_Study}
\begin{figure*}[htpb]
    \centering
    \includegraphics[width=\textwidth]{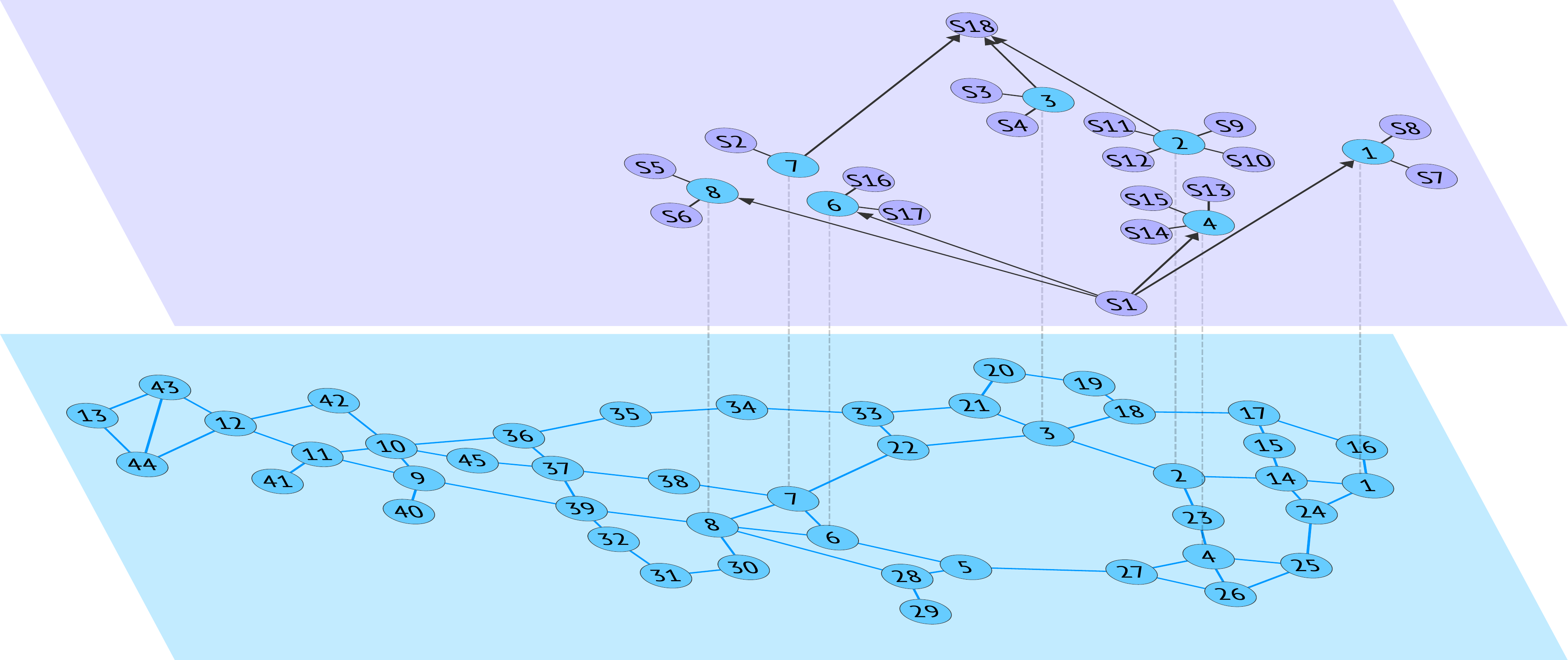}
    \caption{
    The combined graph used in the experiments. The bottom layer represents the original communication network graph and the top layer represents the attached attack states connected to the relevant nodes of the network.
    }
    \label{fig:network_graph}
\end{figure*}

To ground the analysis in a realistic rather than synthetic network structure, we chose a formerly operational fiber-optic based network topology with partial radio link redundancy.
The topology, shown in the bottom layer of Fig.~\ref{fig:network_graph}, represents a critical infrastructure backbone covering a large geographic area.
Although we chose a real-world network, any network of similar scale and structure would have been suitable for demonstrating the applicability of our modelling approach.

To simulate possible attacks on the services of this network, we used an attack graph that represents vulnerabilities found in a multi-cloud enterprise network originally presented in~\cite{stergiopoulos2022automatic}.
Although the vulnerabilities of the attack graph are no longer relevant in today's cybersecurity context, our model itself remains valid as, e.g. CVSS
has always derived exploitability and impact scores for the vulnerabilities.
Most importantly, using vulnerabilities already analysed and presented in the literature makes it easier to compare the results.

To combine the attack graph with the network graph, as a first step, we grouped the attack states by the affected services as presented in \cite{stergiopoulos2022automatic}.
For each of the services, we selected a representative node in the network graph. This mimics the situation in which the software is installed in the network's systems (nodes), and an expert creates the attack graph based on the network's vulnerabilities. The grouping of attack states in services and mapping of services to the selected network nodes is listed in Table~\ref{tab:grouping_attack_states} and is shown in the top layer in Fig.~\ref{fig:network_graph}.

\begin{figure}[htpb]
    \centering
    \includegraphics[width=\columnwidth]{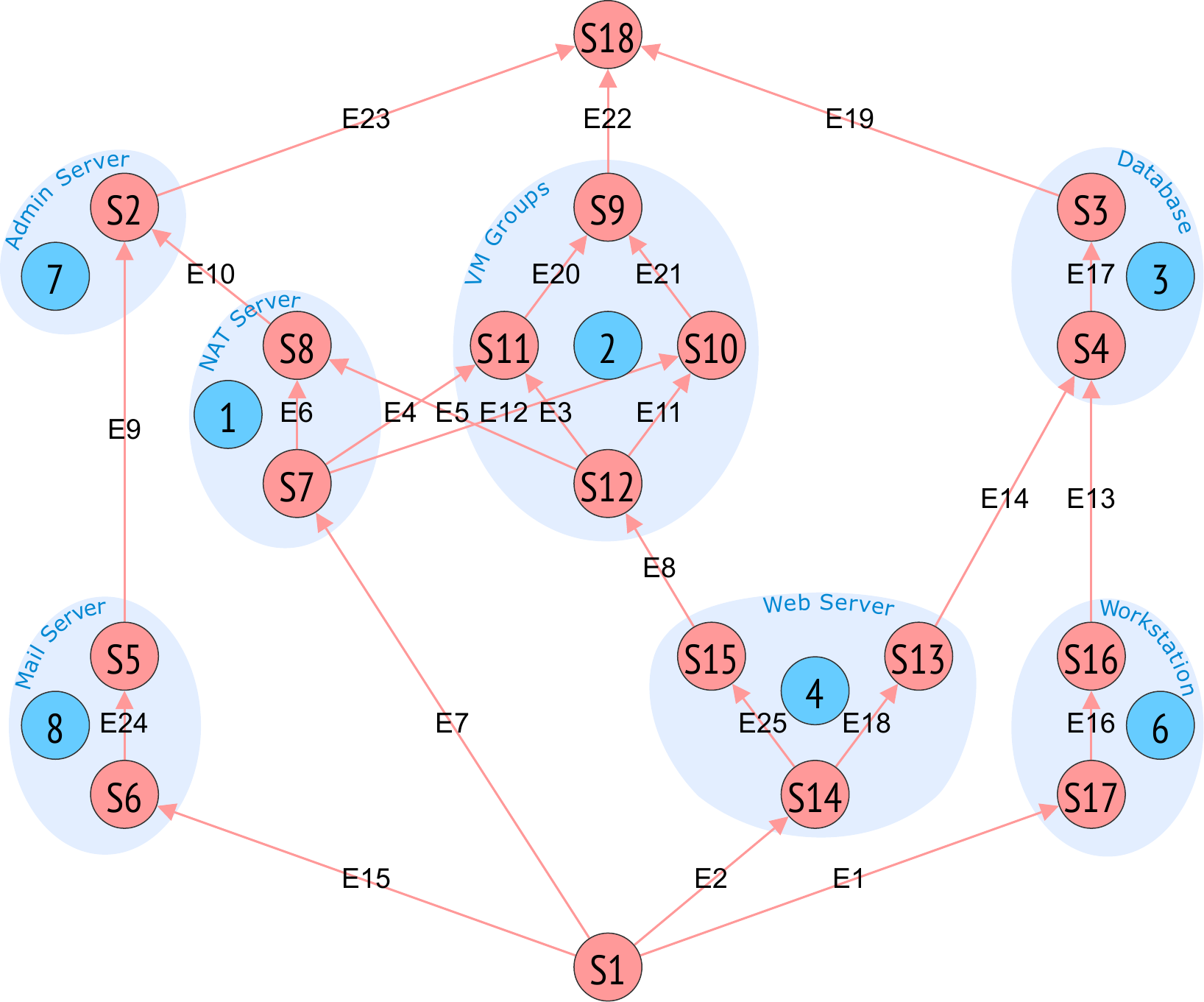}
    \caption{Attack graph used in the experiments with the states grouped by affected services. $S1$ and $S18$ are the start and end states, respectively.}
    \label{fig:attack_graph}
\end{figure}

The attack graph, as well as the grouping of attack states in the network nodes, are depicted in Fig.~\ref{fig:attack_graph}.
The attack graph consists of 18 states (start, end, and 16 attack states), 9 possible attack paths from start to end, and a total of 14 distinct exploitable vulnerabilities. Each vulnerability is represented by one or several directed edges in the attack graph and has both an impact value and an exploitability value calculated using CVSS. The computational methodology for these values is presented in~\cite{kuikka2024networkmodelling} and follows the guidelines shown in~\cite{Scoring}.
The list of vulnerabilities, their impact values, exploitability, and their corresponding edges in the attack graph are presented in Table~\ref{tab:vulnerabilities}.

\begin{table}[htpb]
    \centering
    \caption{Attack states grouped by their affected service and associated network node.\label{tab:grouping_attack_states}}
    \begin{tabular}{|l|l|l|}
    \hline
    \textbf{Network Service} & \textbf{Node No.} & \textbf{Attack States} \\ \hline
    NAT Server               & 1                    & S7, S8                 \\ \hline
    VM Groups                & 2                    & S9, S10, S11, S12      \\ \hline
    Database                 & 3                    & S3, S4                 \\ \hline
    Web Server               & 4                    & S13, S14, S15          \\ \hline
    Workstation              & 6                    & S16, S17               \\ \hline
    Admin Server             & 7                    & S2                     \\ \hline
    Mail Server              & 8                    & S5, S6                 \\ \hline
    \end{tabular}
\end{table}

\begin{table*}[htpb]
    \centering
    \caption{Vulnerabilities present in the attack graph originally presented in~\cite{kuikka2024networkmodelling}.\label{tab:vulnerabilities}}
    \begin{tabular}{|l|l|l|l|}
    \hline
    \textbf{Vulnerability} & \textbf{Exploitability / 10} & \textbf{Impact} & \textbf{Edge(s)}          \\ \hline
    MALICIOUS-WEBSITE      & 1.0                          & 0.0             & E1                      \\ \hline
    DIRECT-ACCESS          & 1.0                          & 0.0             & E2                      \\ \hline
    CVE-2010-3847          & 0.339258                     & 10.00085        & E3, E4, E20             \\ \hline
    CVE-2003-0693          & 0.99968                      & 10.00085        & E5, E6                  \\ \hline
    CVE-2007-4752          & 0.99968                      & 6.442977        & E7                      \\ \hline
    CVE-2001-0439          & 0.99968                      & 6.442977        & E8                      \\ \hline
    CVE-2008-4050          & 0.85888                      & 10.00085        & E9, E10                 \\ \hline
    CVE-2008-0015          & 0.85888                      & 10.00085        & E11, E12, E21, E22, E23 \\ \hline
    CVE-2009-1918          & 0.99968                      & 10.00085        & E13, E16                \\ \hline
    CVE-2018-7841          & 0.99968                      & 6.442977        & E14, E18                \\ \hline
    CVE-2004-0840          & 0.99968                      & 10.00085        & E15                     \\ \hline
    CVE-2008-5416          & 0.7952                       & 10.00085        & E17, E19                \\ \hline
    CVE-2001-1030          & 0.99968                      & 6.442977        & E24                     \\ \hline
    CVE-2009-1535          & 0.99968                      & 6.442977        & E25                     \\ \hline
    \end{tabular}
\end{table*}

\subsection{Initial results}\label{subs:baseline}
In our initial analysis, we set all network link weights to $w_n=1.0$ denoting a $100\%$ probability that a communication or an attack will succeed at the network level. This is effectively the same as only considering the attack graph without there being any defense mechanisms present in the network preventing the attack. This is done to get an understanding of how the different cumulative metrics behave without, yet, taking the
network into account.

\begin{figure}[htpb]
    \centering
    \includegraphics[width=.9\columnwidth,trim=1mm 1mm 1mm 1mm,clip]{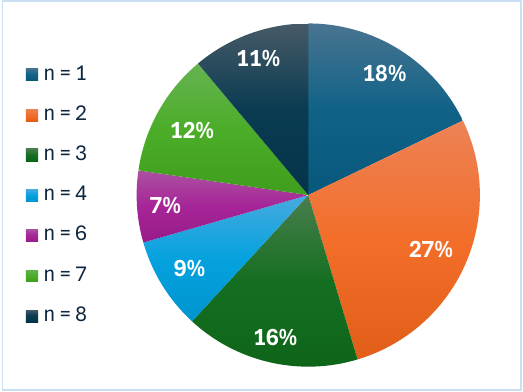}
    \caption{Proportional node-wise cumulative impact $C_{NODE}(n)$ per node $n$, expressed as a fraction of the total, $\sum_{n} C_{NODE}(n)$.}
    \label{fig:init_node_wise}
\end{figure}

In Fig.~\ref{fig:init_node_wise}, we present the node-wise cumulative impact $C_{NODE}$ for all affected services (nodes), expressed as a fraction of the total, $\sum_{n} C_{NODE}(n)$. Here, it is clearly seen that the expected impacts of the vulnerabilities that affect node~2 sum up to be the biggest concern in the attack graph. This is not surprising, since the service in node 2 has the biggest attack surface, i.e., it has the most vulnerabilities providing a foothold to the node and thus the probability of successfully attacking this node is expected to be high. The second explaining factor is that the service in node 2 has the largest overall number of vulnerabilities, so there is a possibility of a higher impact to begin with even if we take into account that it has the least exploitable vulnerabilities $E3$, $E4$, and $E20$ as seen in Tables~\ref{tab:grouping_attack_states} and~\ref{tab:vulnerabilities}.
The $C_{NODE}$ values of the rest of the nodes are broadly consistent with this reasoning, as nodes 1, 3, and 7 have the next highest number of affecting vulnerabilities and the next largest attack surfaces, which is indicated by their high $C_{NODE}$ values, whereas $C_{NODE}(6)$ and $C_{NODE}(8)$ are significantly lower.
The low $C_{NODE}(4)$ value, although more vulnerabilities affect node~4 than node 8, can be explained by the fact that the vulnerability $E2$ does not have an impact.

\begin{figure}[htpb]
    \begin{minipage}{\columnwidth}
        \centering
        \begin{minipage}{\textwidth}
            \subfloat[]{
                \includegraphics[width=\textwidth]{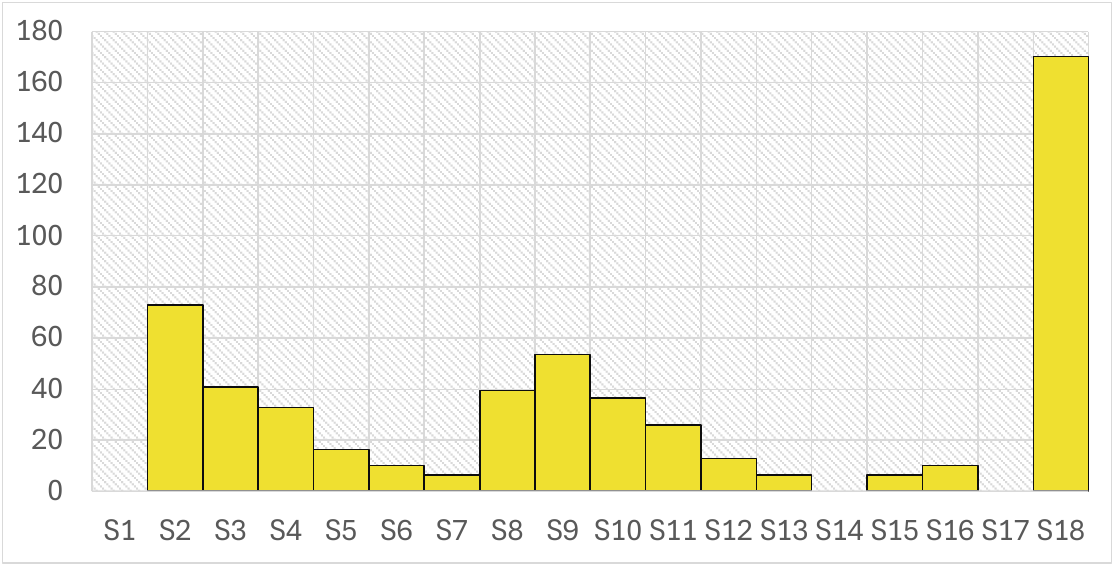}
                \label{fig:init_att_state_wise}
            }
        \end{minipage}
        \begin{minipage}{\textwidth}
            \subfloat[]{
                \includegraphics[width=\textwidth]{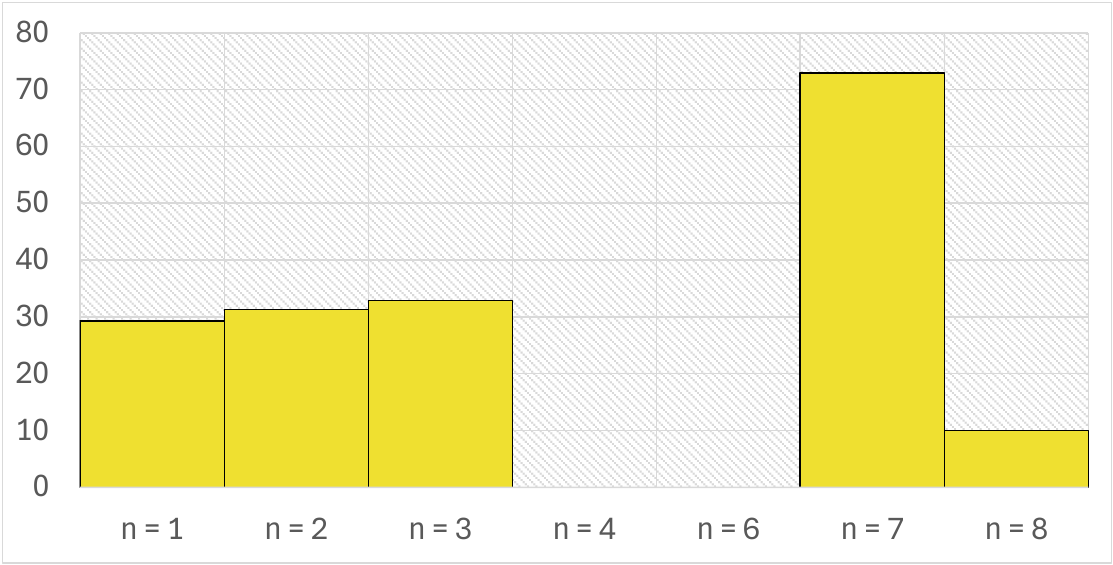}
                \label{fig:init_inbound}
            }
        \end{minipage}
        \begin{minipage}{\textwidth}
            \subfloat[]{
                \includegraphics[width=\textwidth]{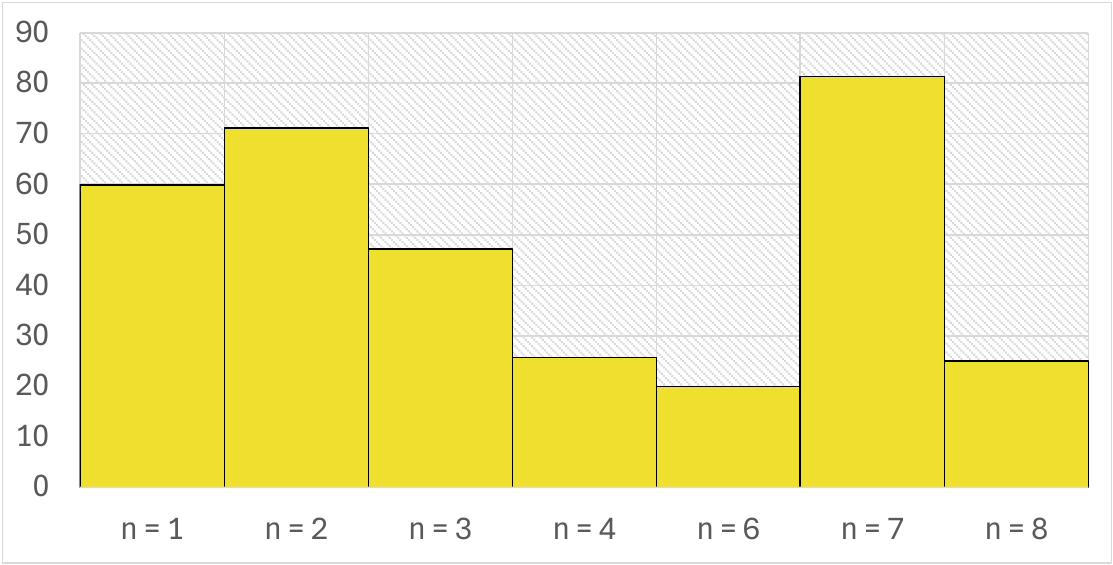}
                \label{fig:init_outbound}
            }
        \end{minipage}
        \caption[]{
        Initial results for expected cumulative impact (vertical axis) across different metrics.
        \subref{fig:init_att_state_wise} $CE[I_i]$, expected cumulative impact per attack state~\eqref{eq:CEI}.
        \subref{fig:init_inbound} $C_{IN}(n)$, expected cumulative inbound impact per node~\eqref{eq:ICEI}.
        \subref{fig:init_outbound} $C_{OUT}(n)$, expected cumulative outbound impact per node~\eqref{eq:OCEI}.
        }
    \end{minipage}
\end{figure}

The results of the expected cumulative impact of each of the attack states ($CE[I]$, see~\eqref{eq:CEI}) are shown in Fig.~\ref{fig:init_att_state_wise}. As expected, the latter states of the attack score higher than the earlier states of the attack (e.g. $CE[I_{S11}] > CE[I_{S15}]$ and $CE[I_{S3}] > CE[I_{S15}]$)
and the cumulation of expected impact within a service can be witnessed (e.g. $CE[I_{S12}]<CE[I_{S10}]<CE[I_{S9}]$).
In particular, $S2$ has a very high expected cumulative impact due to many attack paths consisting of vulnerabilities with high exploitability and a high impact that leads to it. Note that at the beginning of an attack ($S1$), there is no cumulative impact ($CE[I_{S1}]=0$) since the attack has not yet started. Furthermore, since the attack states $S14$ and $S17$ are only reached from $S1$ by exploiting vulnerabilities that do not have impact ($E2$ and $E1$, respectively), the expected cumulative impact of reaching these states is also nonexistent. Since $S18$ is the state in which an attacker reaches its objectives, $CE[I_{S18}]$ denotes the expected cumulative impact of a successful attack ($S1{\rightarrow}S18$). As all the attack paths lead to $S18$, this value consists of the expected cumulated impact of all the attack paths, and therefore the score is higher than for the rest of the attack states. Logically, the cumulative impact is highest at the end of an attack.

A similar trend is reflected in the expected cumulative inbound ($C_{IN}$,~\eqref{eq:ICEI}) and cumulative outbound ($C_{OUT}$,~\eqref{eq:OCEI})
impact metrics as can be seen in Fig.~\ref{fig:init_inbound}~and~\ref{fig:init_outbound}, respectively.
Measured by $C_{IN}$,
the nodes 1, 4, 6 and 8, which provide a foothold in the network, score lower than the nodes further down on the attack path. This is also the general trend with $C_{OUT}$, with the exception of node~1, since clearly $C_{OUT}(3)$ is less than $C_{OUT}(1)$. If we consider these measures to reflect on how large the overall impact is if a node is \emph{attacked} ($C_{IN}$) and \emph{used to attack other services} ($C_{OUT}$), the reasoning for this is straightforward, since there is only one outbound vulnerability that can be exploited from node 3 while there are three of those that can be exploited from node 1.

What is most noticeable are the very high $C_{IN}(7)$ and $C_{OUT}(7)$ values, even though node~7 has one of the smallest $C_{NODE}$ values. This, again, is due to the nature of the metrics used here. That is, while an attack to the node itself might not be of high concern as it has a low impact, the fact that the attack has propagated that far should be alarming, as the attack has, with a high probability, propagated through services that cause the attack to have a high impact.
Note that $C_{IN}(4)=0$ and $C_{IN}(6)=0$ are special cases due to the fact that the vulnerabilities $E1$ and $E2$ have no impact.

\subsection{Mitigation of individual vulnerabilities}\label{subs:mitigate_vulnerabilities}
In enterprise networks with vulnerabilities of varying degrees, it is important to understand the
impact of each vulnerability to be mitigated or fixed.
In an attack graph, completely fixing a vulnerability can be simulated by setting the exploitability of the affected edges to zero ($p_E~=~0.0$).

By setting the exploitability of a vulnerability to zero and rerunning the analysis, we can effectively measure the importance of an individual vulnerability to the cyberattack.
This analysis can be used to prioritize the mitigation of the vulnerabilities in the network.
As a demonstration, we performed a series of calculations, each after having simulated fixing one vulnerability.
In each run, we calculated the $C_{IN}$, $C_{OUT}$ and $CE[I_{S18}]$ and compared the results with those of the original attack graph.
Here, the change signifies how much the expected cumulative impact changed on existing attack paths from the beginning of an attack ($S1$) to the relevant attack states.
Since fixing a vulnerability results in at least one attack path having a vulnerability with zero exploitability ($p_E = 0.0$), thus making the path infeasible, the impact was calculated only for the portion of the path that remains feasible.

\begin{figure*}[htpb]
    \centering
    \includegraphics[width=\textwidth]{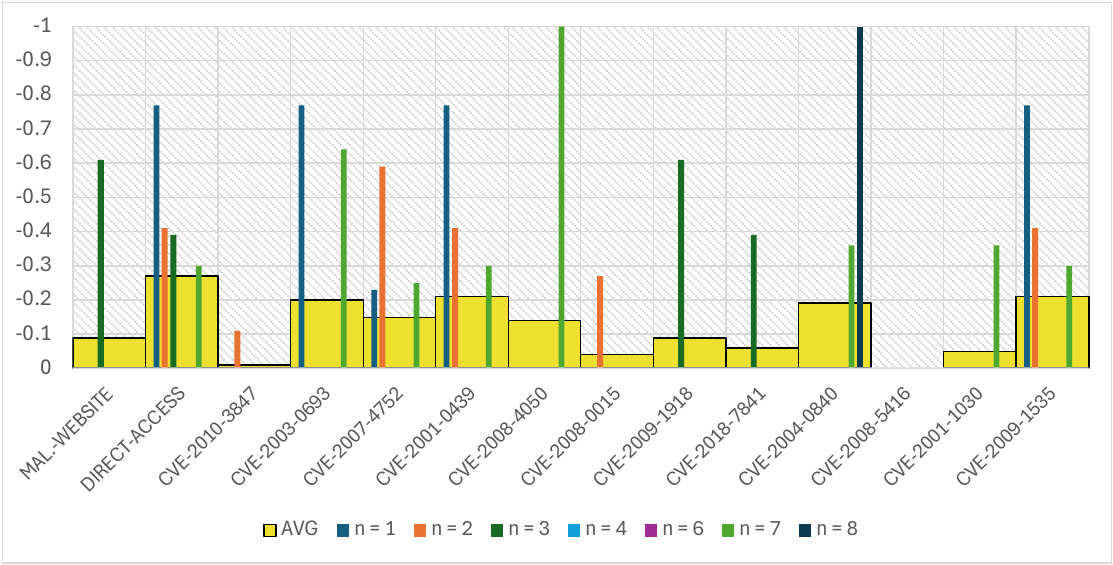}
    \caption{
        Relative change in the node-wise expected cumulative inbound impact, $C_{IN}$, after fixing a vulnerability.
        The individual bars represent the relative change of $C_{IN}$ per node~$n$. Note the reversed vertical axis.
    }
    \label{fig:rem_vuln_nodes_inbound_delta}
\end{figure*}

Fig.~\ref{fig:rem_vuln_nodes_inbound_delta} depicts the relative changes of node-wise expected cumulative inbound impacts ($C_{IN}$) as well as their averages after mitigating a vulnerability. Intuitively, one could assume that fixing the vulnerabilities that work as entry points into the network would have the biggest effect on the cumulative impact, as fixing these vulnerabilities often makes multiple attack paths impossible.
This intuition is mainly supported by our results, as fixing vulnerabilities that provide a foothold in the network seems to have the largest overall effect on $C_{IN}$. For example, \emph{DIRECT-ACCESS} has, on average, the highest relative change as measured by $C_{IN}$, whereas vulnerabilities that are exploited later in the attack, such as \emph{CVE-2018-7841}, have a significantly lower relative change. Since initially both $C_{IN}(4)=0$ and $C_{IN}(6)=0$, the relative change is also zero for these nodes, even when it is impossible to attack them after fixing the vulnerabilities that provide a foothold in these nodes.

In some cases, fixing a vulnerability in the early stages of the attack completely eliminates all $C_{IN}$ for a certain node, while in other cases several vulnerabilities need to be fixed to fully mitigate all $C_{IN}$ to a node. For example, by mitigating \emph{CVE-2004-0840}, all $C_{IN}(8)$ is eliminated, but achieving the same for node~2 requires fixing both \emph{CVE-2007-4752} and \emph{CVE-2009-1535}.

Since we measure the change of cumulative impact on the full existing attack path, we can see that fixing a vulnerability does not, in general, fix the whole attack path to a node. However, the earlier the attack path is mitigated, the bigger the change in cumulated impact. In Fig.~\ref{fig:rem_vuln_nodes_inbound_delta}, this effect can be seen by examining the
relative change of $C_{IN}(1)$ on the attack path consisting of vulnerabilities \emph{CVE-2009-1535}, \emph{CVE-2001-0439} and \emph{CVE-2003-0693} ($E25$-$E8$-$E5$ in Fig.~\ref{fig:attack_graph}).
Clearly, the change is larger if \emph{CVE-2009-1535} is mitigated instead of \emph{CVE-2003-0693}.

The relative change in the expected cumulative outbound impact ($C_{OUT}$) can be seen in Fig.~\ref{fig:rem_vuln_nodes_inbound_delta}.
These results are consistent with those of the
$C_{IN}$ case. That is, fixing vulnerabilities in the early states of the attack path has a larger change than fixing vulnerabilities in the later states.
Also, as expected, vulnerability \emph{CVE-2009-1535} is one of the most significant ones, as it is part of several attack paths with a high expected cumulative impact.

\begin{figure*}[htpb]
    \centering
    \includegraphics[width=\textwidth]{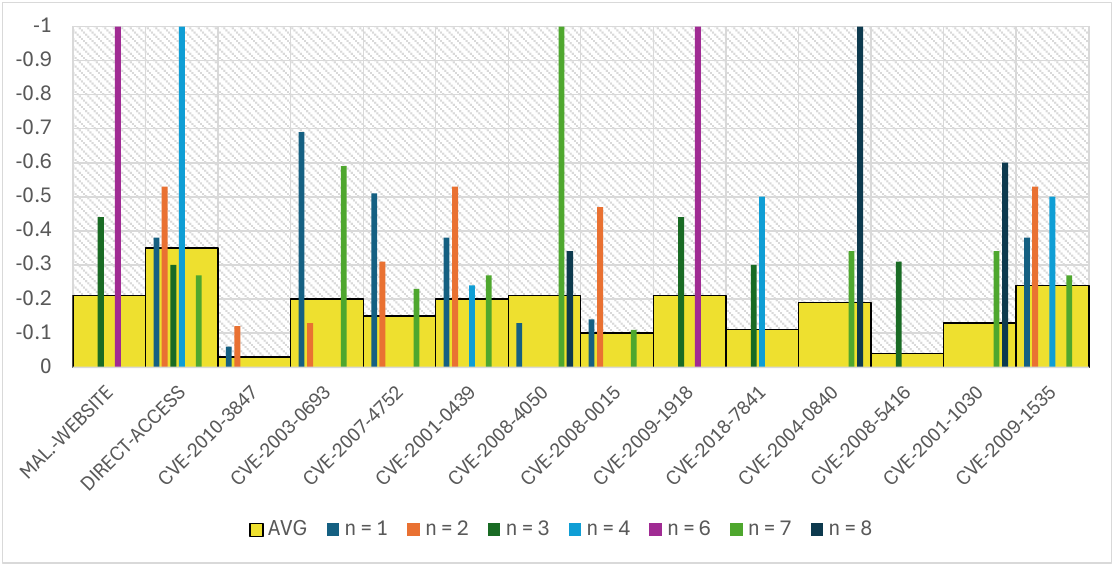}
    \caption{
        Relative change in the node-wise expected cumulative outbound impact, $C_{OUT}$, after fixing a vulnerability.
        The individual bars represent the relative change of $C_{OUT}$ per node~$n$. Note the reversed vertical axis.
    }
    \label{fig:rem_vuln_nodes_outbound_delta}
\end{figure*}

The noticeable difference from the $C_{IN}$ case is the presence of change in nodes~4 and 6.
As mentioned above, removing the vulnerabilities that provide a foothold to these nodes would make it impossible to attack them. This can be seen in Fig.~\ref{fig:rem_vuln_nodes_outbound_delta} by looking at the relative change of
$C_{OUT}(4)$ and $C_{OUT}(6)$
when fixing vulnerabilities \emph{DIRECT-ACCESS} and \emph{MALICIOUS-WEBSITE}, respectively.

Similarly to the $C_{IN}$ case, fixing \emph{CVE-2004-0840} mitigates all $C_{OUT}(8)$
and fixing both \emph{CVE-2007-4752} and \emph{CVE-2009-1535} would eliminate all $C_{OUT}(2)$. Although the former can be directly interpreted on the basis of Fig.~\ref{fig:rem_vuln_nodes_outbound_delta}, the latter can not, as the relative changes of
$C_{OUT}(2)$ after fixing those two vulnerabilities do not sum up to $-1$.
This is due to the fact that while fixing \emph{CVE-2007-4752} blocks all attack paths targeting node~2 through node~1 and fixing \emph{CVE-2009-1535} blocks those that pass through node~4, preventing the attack from progressing through node~2 requires fixing both of them.

\begin{figure}[htpb]
    \centering
    \includegraphics[width=\columnwidth]{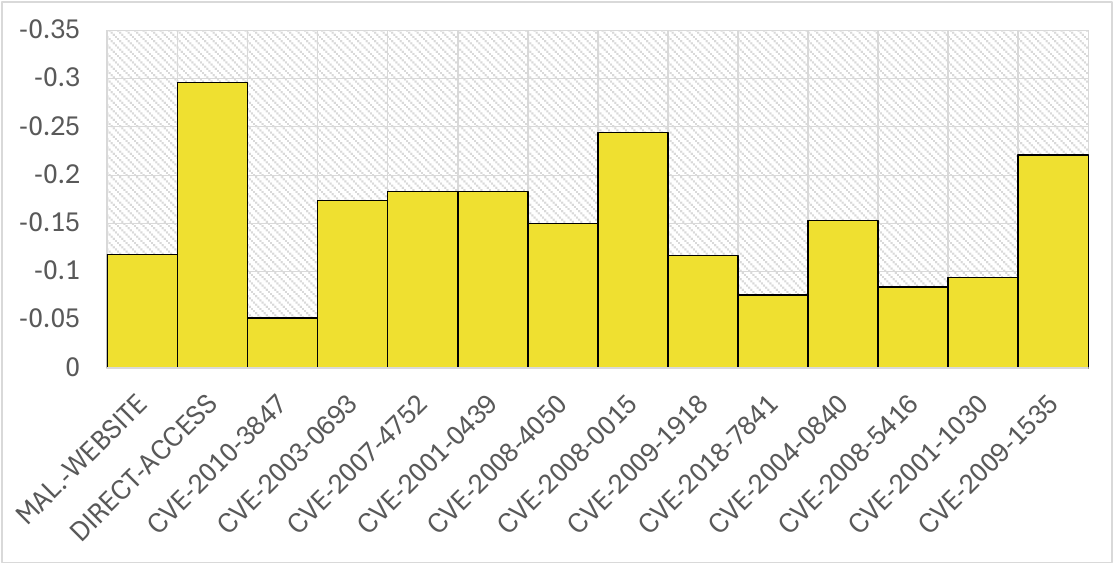}
    \caption{Relative change in the cumulative impact of a successful attack ($CE[I_{S18}]$) after fixing a vulnerability. Note the reversed vertical axis.}
    \label{fig:rem_vuln_overall_delta}
\end{figure}

The relative change in the expected cumulative impact ($CE[I_{S18}]$) of a successful attack ($S1~\rightarrow~S18$) can be seen in Fig.~\ref{fig:rem_vuln_overall_delta}.
In line with the previous findings, fixing vulnerabilities in earlier states has a high relative change. Especially, \emph{DIRECT-ACCESS} is clearly the most impactful vulnerability. In addition and consistent with previous findings, fixing \emph{CVE-2009-1535} results in a high relative change of $CE[I_{S18}]$. An interesting finding is that the vulnerability with the second highest relative change is \emph{CVE-2008-0015}, which is not an entry-point vulnerability and does not have a high relative change as measured by $C_{IN}$ or $C_{OUT}$. One reason for this is the high impact and high exploitability of the vulnerability.
However, the main reason is the number of attack paths that exploit this vulnerability, which can be seen by locating the numerous edges of the vulnerability (Table~\ref{tab:vulnerabilities}) on the attack graph (Fig.~\ref{fig:attack_graph}).

To summarize, we can identify two different types of high-impact vulnerabilities: those that contribute to multiple edges in the attack graph, and those that have edges positioned early in the attack graph.
The vulnerability \emph{CVE-2008-0015} is an example of the former while \emph{DIRECT-ACCESS} and \emph{CVE-2009-1535} are examples of the latter. In both cases, the change in the impact correlates with how many attack paths traverse these edges in general, but also depends on the impacts and probabilities of these paths downstream of the edges. Consistent with this, vulnerabilities on attack paths leading to nodes that have a high $C_{NODE}$ seem to dominate the results. There are also exceptions to the general rule.
For example, \emph{CVE-2010-3847} is a vulnerability of high impact, and it is represented by multiple edges in the attack graph, but due to its low exploitability, it only has a small role in the results.
As another example, \emph{MALICIOUS-WEBSITE} is the first step on an attack path, but because it only affects a single path, fixing it would have a relatively low effect on expected impact.
Importantly, the changes measured by the metrics $CE[I_{S18}]$, $C_{IN}$, and $C_{OUT}$ correlate with each other and help to identify which vulnerabilities should be prioritized for mitigation.

\subsection{Securing individual nodes}\label{subs:node_monitor}
Fixing a vulnerability completely is not always possible for various reasons.
For example, in our scenario, one exploitable vulnerability that cannot be fully fixed is "user browsing a malicious web page".
As a more general example, there may not be a patch available for the vulnerability or the patch cannot be applied due to incompatibility issues with other systems or due to it causing unwanted downtime in operational systems.
In these cases, mitigation strategies include other security measures, such as implementing monitoring solutions to help detect suspicious activity. These measures may not completely stop a cyberattack, but they decrease the probability of its success.

In our analysis, we assumed that mitigation measures represent more robust monitoring of a single node for suspicious network traffic. As the monitoring targets the inter-node traffic, we assume that adding monitoring to a node only affects the exploitability of vulnerabilities that provide a foothold in the node or enable attacks to other nodes in the network while the exploitability of the node's internal vulnerabilities remains unaffected.
Similarly to the vulnerability mitigation case, the analysis can also be used as a measure of the importance of individual nodes in the attack.

To investigate the role of enhanced monitoring of a single node as a mitigation measure, we performed another series of calculations, each time decreasing the probability of a successful attack against or from a node, one node at a time, and comparing the results to those of the original run.
That is, the exploitability of the node's inbound and outbound vulnerabilities was affected by a common factor~$f$:
\begin{equation*}
\hat{p}_{E}(i,j) = f * p_{E}(i,j).
\end{equation*}
The internal vulnerabilities remained unchanged.
In our calculations, we used a factor of $f=0.2$, which means that the added monitoring solution prevents $80\%$ of exploit attempts and, thus, reduces the exploitability of each affected vulnerability by $80\%$.

\begin{figure}[htpb]
    \centering
    \includegraphics[width=\columnwidth]{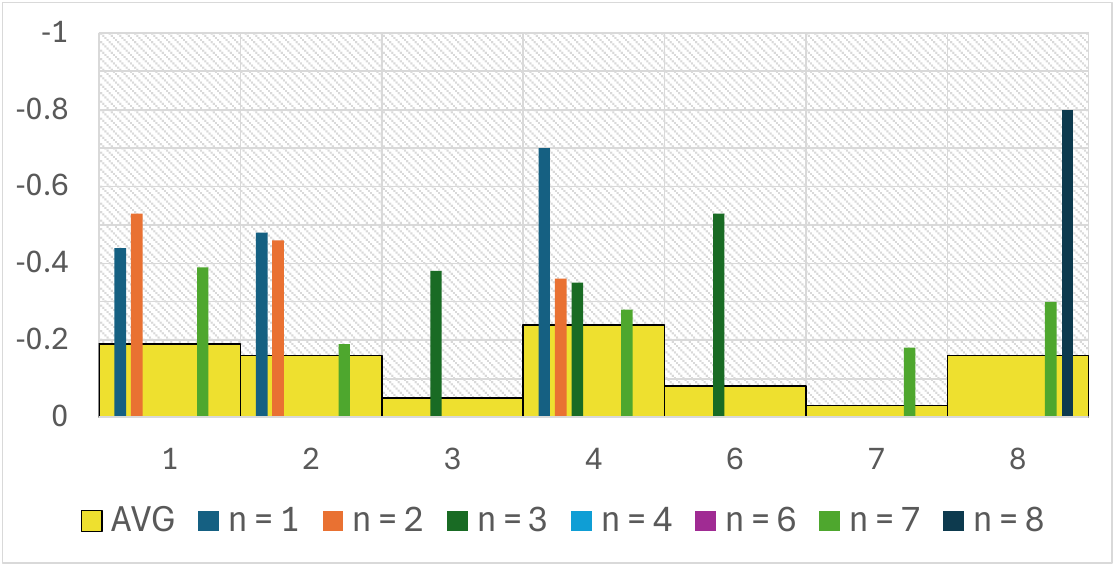}
    \caption{
        Relative change in the node-wise expected cumulative inbound impact, $C_{IN}$, after adding monitoring to a network service. The horizontal axis represent the number of the node with added monitoring. The individual bars represent the relative change of $C_{IN}(n)$ per node~$n$. Note the reversed vertical axis.
        }
    \label{fig:mon_nodes_inbound_delta}
\end{figure}

\begin{figure}[htpb]
    \centering
    \includegraphics[width=\columnwidth]{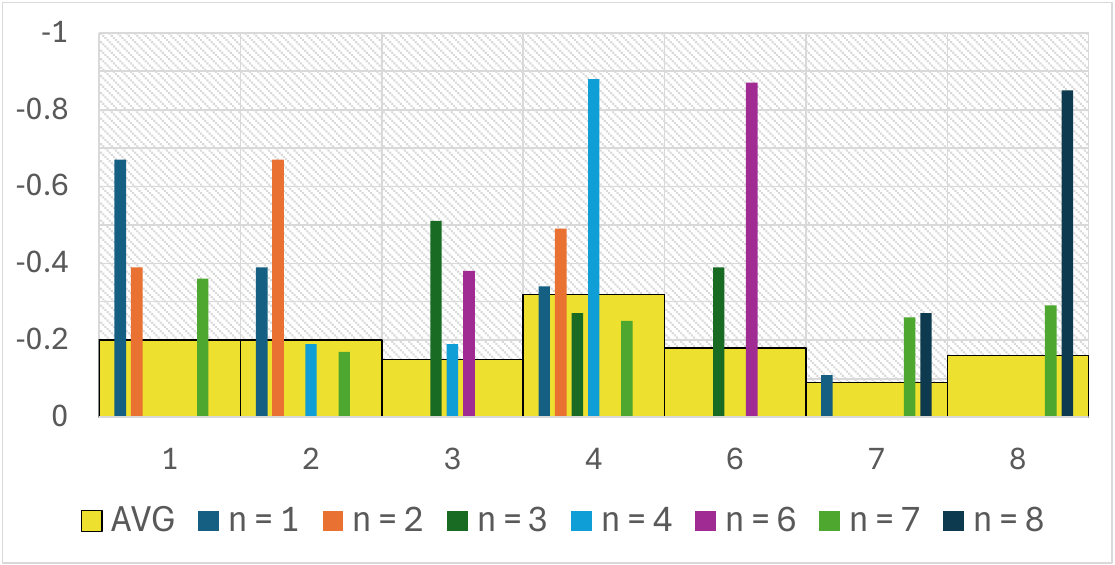}
    \caption{
        Relative change in the node-wise expected cumulative outbound impact, $C_{OUT}$, after adding monitoring to a network service. The horizontal axis represent the number of the node with added monitoring. The individual bars represent the relative change of $C_{OUT}(n)$ per node~$n$. Note the reversed vertical axis.
    }
    \label{fig:mon_nodes_outbound_delta}
\end{figure}

The relative changes of the node-wise expected impacts $C_{IN}$ and $C_{OUT}$ are presented in Fig.~\ref{fig:mon_nodes_inbound_delta}~and~\ref{fig:mon_nodes_outbound_delta}, respectively. The results we obtained are consistent with those acquired from the vulnerability mitigation simulations. It is obvious that adding monitoring to a node mainly affects its own expected cumulative inbound and outbound impacts, with the exception of $C_{IN}(4)$ and $C_{IN}(6)$, which were already zero in the initial case. More interesting is how added monitoring affects the inbound impact of nodes later in the attack path and the outbound impact of nodes both earlier and later in the attack path. In general, adding monitoring to the nodes that are targeted early in the attack (nodes 1, 4, 6, and 8) tends to have higher relative change than adding monitoring to those that are targeted later (i.e., nodes 2, 3, and 7). By adding monitoring to the early states of the attack, the probability of successfully exploiting the early states of the attack without being caught is reduced, and therefore the probability of reaching the latter states is also reduced, which in turn reduces the expected impact of the attack.

Unsurprisingly, adding monitoring to node~4 resulted in the largest relative change in $C_{IN}$ and $C_{OUT}$ when averaged over the nodes. This could be expected, as fixing its foothold vulnerability \emph{DIRECT-ACCESS} resulted in the largest relative change in $C_{IN}$, $C_{OUT}$ and $CE[I_{S18}]$ in the vulnerability fixing case. Also consistent with the previous findings is that adding monitoring to node~2 resulted in high relative changes of both $C_{IN}$ and $C_{OUT}$. This is explained by vulnerabilities affecting node~2, such as \emph{CVE-2008-0015} and \emph{CVE-2001-0439},
both of which have a large change in $CE[I_{S18}]$ in the vulnerability mitigation case.

\begin{figure}[htpb]
    \centering
        \includegraphics[width=\columnwidth]{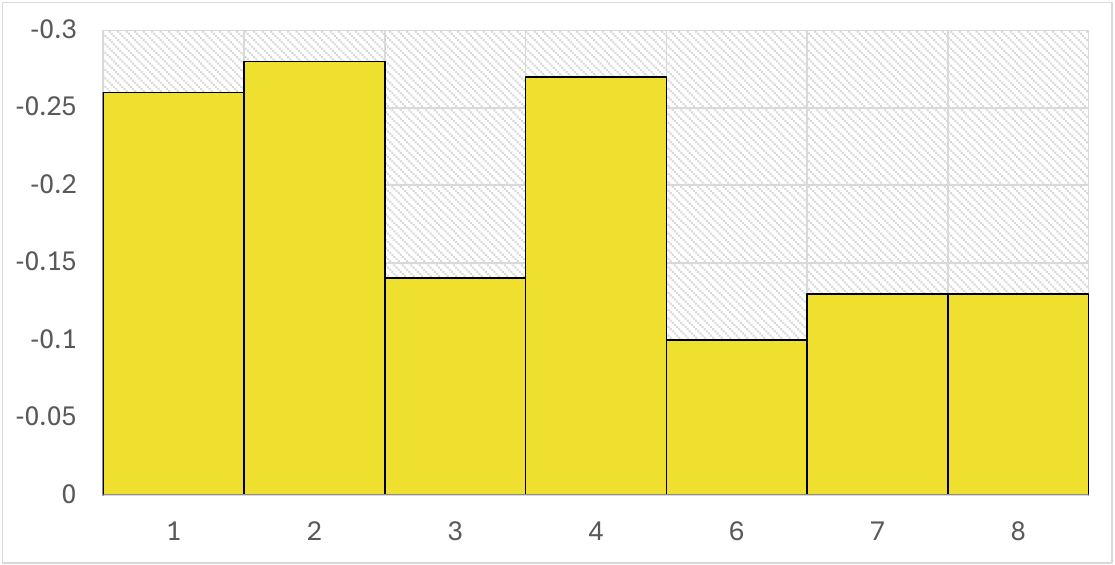}
    \caption{
        Relative change in the cumulative impact of a successful attack ($CE[I_{S18}]$) after adding monitoring to a network service.
        The horizontal axis represent the number of the node with added monitoring.
        Note the reversed vertical axis.
    }
    \label{fig:mon_nodes_overall_delta}
\end{figure}

Fig.~\ref{fig:mon_nodes_overall_delta} shows the relative change in the expected cumulative impact of a successful attack ($CE[I_{S18}]$) after adding monitoring to each of the nodes affected by the attack individually. Although these results mainly conform to the relative change measured by $C_{IN}$ and $C_{OUT}$, there are some differences.
For example, adding monitoring to node~2 reduces the expected cumulative impact the most when analysing the entire attack. While this
does not reflect the $C_{IN}$ and $C_{OUT}$ values, it is consistent with the findings of $CE[I_{S18}]$ in the vulnerability mitigation case (see Fig.~\ref{fig:rem_vuln_overall_delta}), namely high relative changes of vulnerabilities affecting node~2, such as \emph{CVE-2008-0015}.
In addition, this can be further explained by the high number of attack paths that exploit the node.

Another clear difference on the scale of the entire attack is the significance of adding monitoring to node~7 compared to the
$C_{IN}$ and $C_{OUT}$ cases, in which node~7 could be interpreted as the least significant node in terms of relative change in cumulative impact after added monitoring. When analysing the relative change of adding monitoring to it on the scale of the whole attack, it is equally significant to node~8 and more significant than node~6. This can be explained in a similar way as the factor of the importance of vulnerabilities (such as \emph{CVE-2008-4050}) on $CE[I_{S18}]$ affecting node~7 and the number of attack paths exploiting it.

The nodes whose vulnerabilities ranked high in the vulnerability mitigation case also rank high in the node monitoring case.
Notably, nodes~1, 2, and 4 are the most impactful ones, which is explained by their high-scoring vulnerabilities \emph{CVE-2007-4752}, \emph{CVE-2008-0015}, and \emph{DIRECT-ACCESS}, respectively.
In general, the entry point nodes score higher than those later in the attack paths, with the exception of node~2, the high score of which is explained by the high number of attack paths traversing it.
Similarly, the lower scores of entry-point nodes~6 and 8 can be explained by the small number of paths traversing them.
The impact metrics $CE[I_{S18}]$, $C_{IN}$, and $C_{OUT}$ correlate with each other for the most part.
The exception is node~2, where the high expected cumulative impact of a successful attack, $CE[I_{S18}]$, reduction cannot be directly seen in the relative changes of expected cumulative inbound and outbound impacts, $C_{IN}$ and $C_{OUT}$, respectively.
It is, however, consistent with the results of Section~\ref{subs:mitigate_vulnerabilities} and helps prioritize which nodes to monitor.
When patching a vulnerability is not possible, reinforced monitoring of the affected node seems to offer similar results, although full mitigation is not achieved.

\subsection{Varying network link weights}\label{subs:link_weights}
So far, we have maintained the weights of the network links at $w_n=1.0$, which means that the state-to-state probabilities have been $p_N=1.0$ at the network level. In practice, this means that if the attacker has sufficient skills or other resources (the amount represented by "exploitability"), the attack is always successful. In the attack mitigation and node monitoring scenarios, we only made changes to the relevant vulnerability weights as both scenarios only affected the exploitability of the known vulnerabilities of the services.
In the present scenario, we demonstrate how the metrics behave if the underlying weights of the network link weights ($w_n$) are varied.
In real-life use cases, this could be analogous to, for example, adding monitoring of suspicious activity to the entire network.

We performed a series of calculations, each after setting all network link weights to a fixed uniform value $w_{n} \in \{0.2, 0.4, 0.6, 0.8, 1.0\}$, to systematically sweep the full probability range.
The network graph with newly set link weights was combined with the attack graph, and the overall state-to-state probabilities were calculated. Since link weights only affect node-to-node traffic, the probability of successful exploitation of a vulnerability is unaffected when the $w_n$ is decreased if the source and target states of a vulnerability are in the same node. If, however, they are not, the more node-to-node hops there are between the states, the more the possibility of a successful exploitation decreases due to the $p_N$ in~\eqref{eq:prob_ovr} being dependent on the underlying network link weights. In addition, the fewer possible paths through the network between states, the smaller $p_N$ is.
Therefore, in theory, low network link weights favor attack paths that have fewer node-to-node hops and more possible paths between the attack states, but this is not always evident from our results, as our metrics do not focus on individual attack paths.

\begin{figure}[htpb]
    \begin{minipage}{\columnwidth}
        \begin{minipage}{\textwidth}
            \subfloat[]{
                \includegraphics[width=\textwidth]{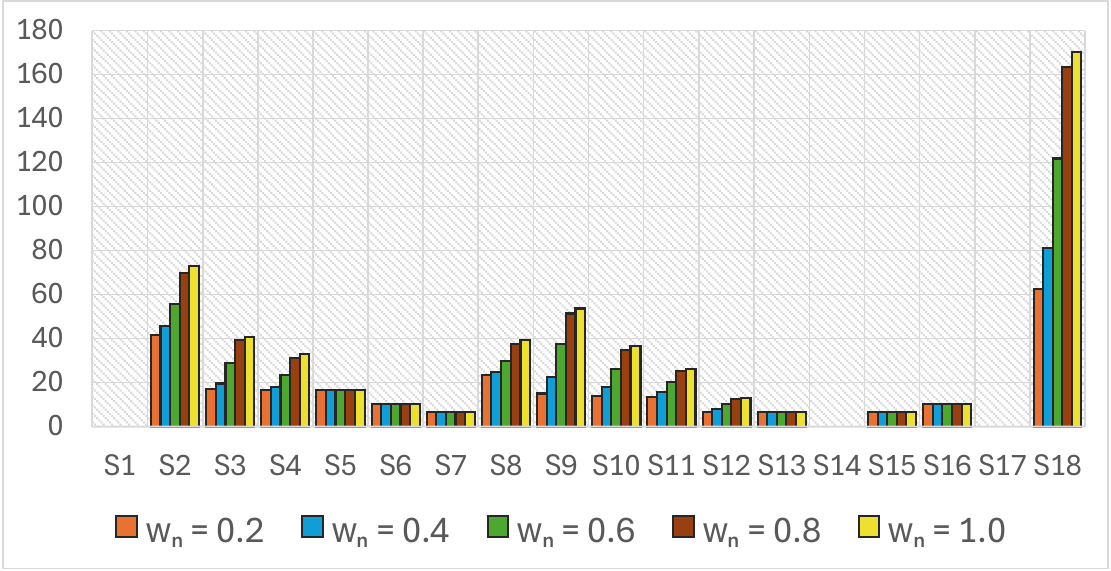}
                \label{sfig:vary_init_attack}
            }
        \end{minipage}
        \begin{minipage}{\textwidth}
            \subfloat[]{
                \includegraphics[width=\textwidth]{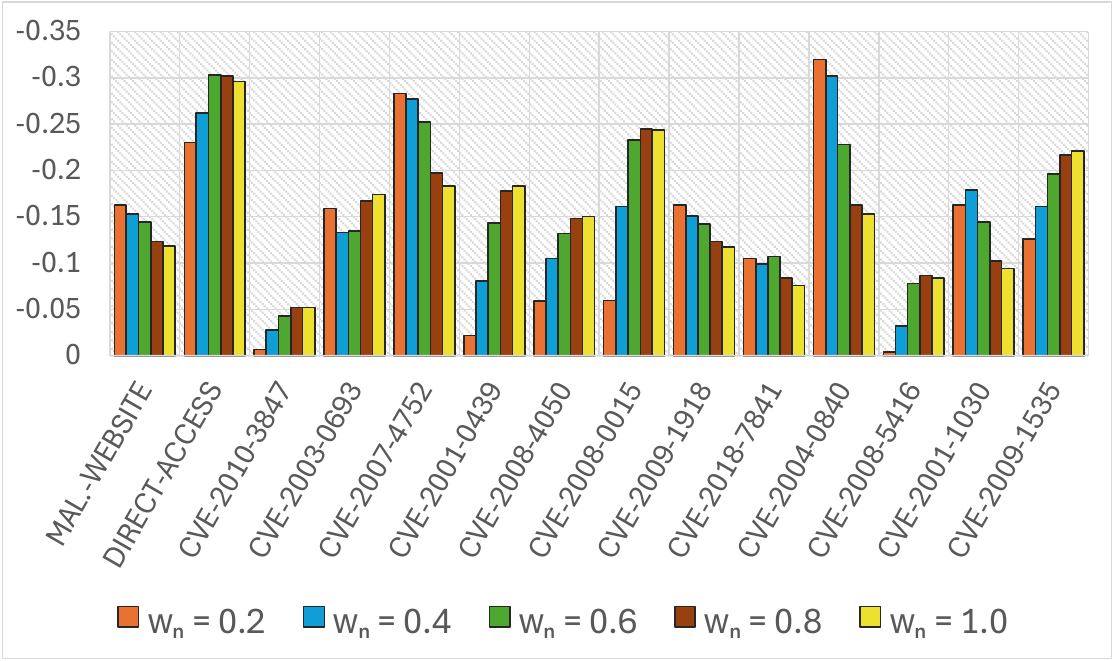}
                \label{sfig:vary_rem_vuln_end}
            }
        \end{minipage}
        \begin{minipage}{\textwidth}
            \subfloat[]{
                \includegraphics[width=\textwidth]{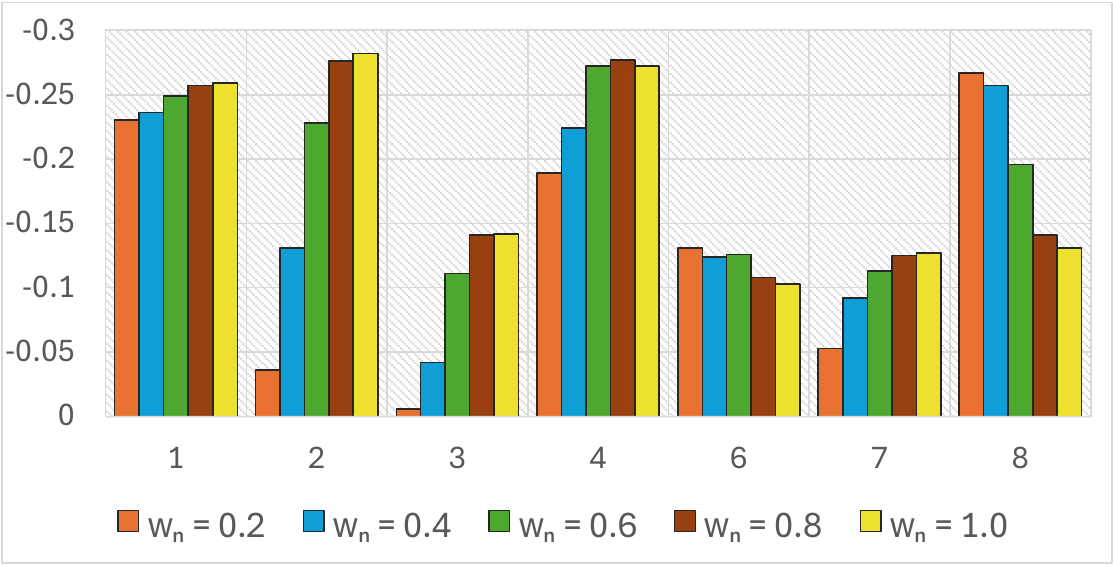}
                \label{sfig:vary_monitor_end}
            }
        \end{minipage}
    \end{minipage}

    \caption[]{
        Influence of varying the network link weights on different impact metrics.
        In each of the figures, the coloured bars represent the values acquired by setting weight of each network link to a fixed value between $0.2$ and $1.0$.
        \subref{sfig:vary_init_attack} Attack state-wise expected cumulative impact, $CE[I_{i}]$.
        \subref{sfig:vary_rem_vuln_end} Relative change in the expected cumulative impact of a successful attack ($CE[I_{S18}]$) after fixing a single vulnerability. Note the reversed vertical axis.
        \subref{sfig:vary_monitor_end} Relative change in the expected cumulative impact of a successful attack ($CE[I_{S18}]$) after adding monitoring to a network service. Note the reversed vertical axis.
    }
    \label{fig:vary_nw_weights}
\end{figure}

The attack state-wise expected cumulative impact ($CE[I]$) is presented in Fig.~\ref{sfig:vary_init_attack}.
In general, there is a clear trend that the network link weights of $w_n=0.8$ have a relatively small effect on the cumulative impact compared to the results obtained with the link weights of $w_n=1.0$.
This is explained by the fact that there are multiple possible paths between the network nodes and with a high link weight, the probability of reaching a node from another node still remains relatively high.
When the weights are further decreased, this probability decreases rapidly,
and there is clearly a larger drop in $CE[I]$ with link weights of $w_n=0.6$ and $w_n=0.4$.
The rate of change in probability is again more subtle when the weights are further decreased to $w_n=0.2$, which is indicated by a smaller drop in $CE[I]$.

Since decreasing $w_n$ does not affect the probability of successful exploitation of vulnerabilities that provide access to the network or of vulnerabilities between attack states within the same node, their expected cumulative impact remains unaffected.
This can be seen in Fig.~\ref{sfig:vary_init_attack} by comparing the $CE[I_{S5}]$ and $CE[I_{S6}]$ obtained with different network weights. However, if a state can be reached from both a state affected by link weights and a foothold state within the same node, its $CE[I]$ is affected by the link weights, which is the case for $CE[I_{S8}]$, as $p_N(S12,S8)$ is dependent on $w_n$.

To understand how the link weights affect the impact of vulnerabilities we repeated the analysis of fixing each vulnerability with varying network link weights and calculated the relative changes in the expected cumulative impact of a successful attack ($CE[I_{S18}]$).
As previously established, with low link weights, the lateral movement from a node to another becomes less likely. This is clearly reflected in the results presented in Fig.~\ref{sfig:vary_rem_vuln_end}, where it is seen that lowering $w_n$ results in an emphasis on mitigating vulnerabilities that can be exploited before there is a hop from node to node in the attack path.
For example, if measured by $CE[I_{S18}]$, \emph{CVE-2001-0439} and \emph{CVE-2008-0015} both are among the most important vulnerabilities when $w_n$ is high but amongst the most trivial if $w_n$ is low.

Mitigating vulnerabilities in the early steps of attack path results in higher relative change in $CE[I_{S18}]$ for low $w_n$ and lower change for high $w_n$, but interestingly, not always. For example, in the case of vulnerability \emph{CVE-2018-7841} in node~4, the relative change tends to be greater the lower the $w_n$ but the opposite is true for \emph{CVE-2009-1535}, which is also a node~4 vulnerability. This is due to the low cumulative impact after exploiting \emph{CVE-2009-1535}, which is caused by the decreased probability of attacking node~3 from node~4 ($p(S13,S4)$). The decreased probability itself can be explained by the crucial role the underlying network plays when attacking from one node to another. Specifically, if $w_n<1.0$, the more hops there are between two nodes in the network, the less likely an attack between them will succeed. This is not only seen with \emph{CVE-2009-1535}, but especially so with \emph{CVE-2008-5416}, which requires that the attack has already reached node~3.
As seen in the attack graph in Fig.~\ref{fig:attack_graph}, the only possible ways to attack node~3 are from nodes 4 and 6. However, due to the network structure, which can be seen in Fig.~\ref{fig:network_graph}, node~3 is at minimum three hops away from node~4 or 6, which makes reaching node~3 extremely improbable during an attack, if $w_n$ is low. Fixing a vulnerability beyond this point becomes less significant with low link weights and, if $w_n=0.2$, fixing \emph{CVE-2008-5416} results in a near-zero relative change of $CE[I_{S18}]$, as demonstrated in Fig.~\ref{sfig:vary_rem_vuln_end}.
These results reflect reality, as stricter network monitoring reduces the likelihood of the attack proceeding deep into the network, making mitigation of vulnerabilities in the early states more critical than in the later states.

Analogous to the vulnerability fixing simulation with varying network link weights, we also ran a series of simulations to analyse the role of the network link weights in the scenario, in which we add monitoring to a node for its known vulnerabilities. That is, in each run, we set the network weights to a fixed value, calculated the expected impact of a successful attack and then calculated the relative change of $CE[I_{S18}]$ after adding monitoring for known vulnerabilities to each of the nodes, one at a time.

Similarly to the vulnerability fixing simulation results, with lower link weights, adding monitoring to nodes that are targeted early in the attack tends to have a higher relative change of $CE[I_{S18}]$ than nodes further down the attack paths.
This is clearly seen in Fig.~\ref{sfig:vary_monitor_end} by looking at the relative change of $CE[I_{S18}]$ after adding monitoring to node~8.
Clearly, if the network link weights are lower and the attack from one node to another is less probable, the change is greater than with higher network link weights.
The same effect can be seen in Fig.~\ref{sfig:vary_rem_vuln_end} by looking at the relative change of $CE[I_{S18}]$ after fixing the vulnerability \emph{CVE-2004-0840}, which provides a foothold into node~8.
In contrast, we can see the opposite effect by looking at the results of node~3 that are nearly identical to those of the vulnerability \emph{CVE-2008-5416} in the previous run. The reason for this is that attacking node~3 becomes highly improbable with low network link weights, and thus adding monitoring for attacks exploiting its known vulnerabilities only yields a small relative change of $CE[I_{S18}]$.

One clear takeaway from this scenario is that lowering the chances of an attacker's lateral movement in the network, for example, by adding monitoring across the whole network (in contrast to per-node monitoring of Section~\ref{subs:node_monitor}), reduces the likelihood of the attack propagating deep into the network.
Furthermore, if the weight of the network links $w_n$ is low, stopping the attack early is most effective, as the probability of the attack propagating deeper into the network decreases.
This clearly further emphasizes the role of entry-point nodes and vulnerabilities, and diminishes the role of, for example, node~2 and its vulnerabilities.
Therefore, in a strictly monitored and protected environment, where the attacker is likely to be caught early, the defensive efforts are best aimed at preventing entry into the network.

\subsection{Discussion and comparison with prior work}

Our initial analysis (Section~\ref{subs:baseline}) identified the nodes~1, 2, and 3 as key impact-inflicting nodes due to their high $C_{NODE}$.
When measured with $C_{IN}$ and $C_{OUT}$, we further recognized node~7 as a key indicator node, signifying that an attack reaching this node has likely already inflicted a significant amount of impact.
Finally, with the added per-node monitoring in Section~\ref{subs:node_monitor}, we recognized the key transit nodes~1, 2, and 4, to which priority should be given when coming up with mitigation strategies.
In Section~\ref{subs:mitigate_vulnerabilities} we recognized \emph{DIRECT-ACCESS}, \emph{CVE-2008-0015}, and \emph{CVE-2009-1535} as our top vulnerabilities.
The same attack graph has been analysed in~\cite{stergiopoulos2022automatic} (Section~5.2), allowing a meaningful comparison between the methodologies.

In~\cite{stergiopoulos2022automatic}, the authors rank system states based on the closeness centrality values of minimum and maximum weight arborescence, and list the top derivative dependency paths based on cumulative risk.
Based on the latter, the authors point out that \emph{CVE-2008-0015} and \emph{CVE-2001-0439} are key vulnerabilities.
While our methodology agrees that the former vulnerability is highly influential, in the case of the latter, our model favors stopping the attack one step earlier and mitigating \emph{CVE-2009-1535}, for a greater reduction in cumulative impact.
However, this vulnerability is listed in their arborescence results, as their analysis highlights attack states $S12$, $S15$, and $S7$ along with their respective vulnerabilities \emph{CVE-2001-0439}, \emph{CVE-2009-1535}, and \emph{CVE-2003-0693}.
A subtle difference between the methods here is that our model prefers mitigating \emph{CVE-2007-4752} over \emph{CVE-2003-0693}.
The major distinction between the methods is that~\cite{stergiopoulos2022automatic} does not consider \emph{DIRECT-ACCESS} as a risk, while ours ranks it the highest.
Assuming access to the web server cannot be prevented, then identifying it as a major threat is valuable input for a security expert, as the best defense might be monitoring this entry point.
Although the model presented in~\cite{stergiopoulos2022automatic} does not quantitatively estimate the impact per service, the attack states and vulnerabilities mostly present in nodes~1, 2, and 4 appear in their top results, in line with our identification of these nodes as most critical.
Overall, both methodologies can be considered valid, but the advantage of our method is that it also quantitatively estimates the influence of nodes and takes the effect of network layer into account.
This supports the analysis of mitigation strategies, such as node monitoring and network segmentation.

Our methodology is not without limitations.
One of them is the creation of a realistic attack graph, which is often manual work by an established expert.
Another shortcoming is the dependence of the approach either on the existing scoring systems, such as CVSS, or on a custom-based vulnerability evaluation.
In the case of the former, the validity of expected impact scores and of the mitigation strategy assessment relies on how accurately the vulnerabilities' exploitability and impact values reflect the reality.
Furthermore, zero-day vulnerabilities and, for example, unauthorized access due to configuration issues (such as the use of default administrator credentials) cannot be estimated with established scoring systems.
On the other hand, a custom evaluation could provide better flexibility, accuracy, and realism, but requires thorough analysis by a cybersecurity expert.
Similarly, estimating the network link weights $w_n$ of our model in practice remains an open challenge.

\section{Conclusion}
\label{sec:Conclusion}

In this study, we have presented a novel methodology to model cyberattacks, analyse their paths through exploitable vulnerabilities, and evaluate the impact of cyber exploits on services in an enterprise or infrastructure network.
Our model takes into account interconnected dependencies across different network layers, from the physical layer to the application layer.

We have employed a probabilistic graph-based approach to analyse cyberattack propagation and its effect on user services in communication networks.
As demonstrated through a realistic case study, our holistic approach, which jointly captures vulnerability exploitability probabilities, network reliability, and service-level impact within a unified framework, proved to be versatile, as it can be used to investigate cyberattacks at various levels of detail, and to evaluate different attack scenarios and mitigation measures.
In addition, our method enables the planning and implementation of tailored protective actions and could help security specialists and system administrators make decisions regarding risk mitigation strategies.

Future work will focus on refining the model by improving the estimation of realistic attack probabilities and impact values.
Furthermore, our methodology would benefit from automated attack graph creation utilizing the output of network vulnerability scanners to reduce its dependency on manual work and make it a practical tool for cyber risk analysis in enterprise networks.

\appendices

\section{\break Cumulative Metrics Example}\label{app:metrics}

\begin{figure*}[htpb]
    \begin{minipage}{\textwidth}
    \begin{center}
    \begin{minipage}{.3\textwidth}
    \begin{center}
    \subfloat[]{
        \includegraphics[width=.7\textwidth]{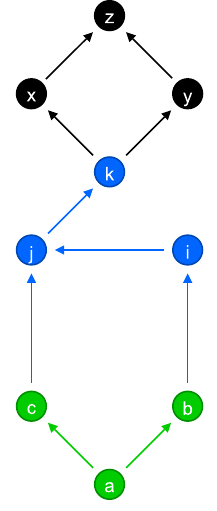}
        \label{sfig:example_graph}
    }
    \end{center}
    \end{minipage}
    \begin{minipage}{.3\textwidth}
    \begin{center}
    \subfloat[]{
        \includegraphics[width=.7\textwidth]{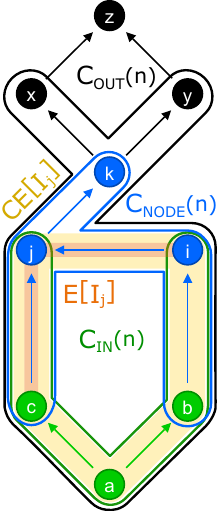}
        \label{sfig:example_metrics}
    }
    \end{center}
    \end{minipage}
    \begin{minipage}{.3\textwidth}
    \begin{center}
    \subfloat[]{
        \includegraphics[width=.7\textwidth]{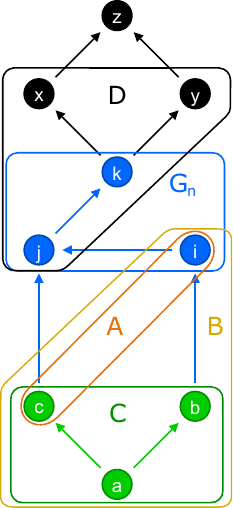}
        \label{sfig:example_sets}
    }
    \end{center}
    \end{minipage}
    \end{center}
    \caption[]{
        Visual representation of the utilized metrics and sets used in their equations. Blue states are grouped to node~$n$ ($G_n$), green and black are the states prior to and after $G_n$ on attack paths, respectively.
        \subref{sfig:example_graph} An attack graph.
        \subref{sfig:example_metrics} Demonstrated metrics. Coloured lines (node~$n$ metrics) and solid colour (state~$j$ metrics) indicate the vulnerabilities, whose impact is included in the corresponding metric.
        \subref{sfig:example_sets} Sets used in the equations of the metrics. Coloured lines indicate the set membership of the attack states. The set's colour matches the metric it belongs to.
    }
    \label{fig:example_metrics_groups}
    \end{minipage}
\end{figure*}

An example of a simple attack graph is presented in Fig.~\ref{sfig:example_graph}.
It consists of nine attack states with ten vulnerabilities between them. The states $a$ and $z$ are the start and end states, respectively.
In this example, the states $i$, $j$ and $k$ are grouped to a node $n$ ($G_n$), as indicated by their blue color.
The green states $a$, $b$ and $c$ precede the $G_n$ states on the attack paths, and the black states $x$, $y$, and $z$ come after the $G_n$ states.

Using node~$n$ and the attack state $j$ as examples, we show which vulnerability impact values are relevant for node-wise metrics $C_{IN}(n)$, $C_{OUT}(n)$ and $C_{NODE}(n)$ (\eqref{eq:ICEI},~\eqref{eq:OCEI} and~\eqref{eq:NWCEI}, respectively), as well as $E[I_j]$ and $CE[I_j]$ (\eqref{eq:E_IMP}~and~\eqref{eq:CEI}, respectively) for attack state~$j$.
A visual representation of these metrics and their included vulnerabilities is shown in Fig.~\ref{sfig:example_metrics}.
To further help explain the metrics, the different sets used in their equations are presented in Fig.~\ref{sfig:example_sets}.
In the figures, the colours of the metrics and the colours of the sets in their equations match, although $G_n$ also appears in the equations for $C_{IN}(n)$ and $C_{OUT}(n)$.

The set $A$ is an example of which states are considered to be \emph{immediately} preceding the state $j$ when calculating $E[j]$ and $B$ is an example of states that are considered to precede the state $j$ on every possible attack path when calculating $CE[I_j]$.
The states grouped to node~$n$ are represented in set $G_n$. All the non-$G_n$ states preceding them on any path and any state that can be attacked from them are represented by sets $C$ and $D$, respectively.

Since the vulnerabilities ($v_{start,end}$) $v_{b,i}$ and $v_{c,j}$ provide a foothold into the node~$n$, their impact values and those of the vulnerabilities preceding them are included in $C_{IN}(n)$. The impact of $v(i,j)$ is excluded as this exploit takes place after gaining access to the node. Similarly, for $C_{OUT}(n)$, the impact values of all the vulnerabilities until states $x$ and $y$ are taken into account. The state $z$ does not immediately follow any $G_n$ state, and is excluded from the set $C$. Thus, the vectors $v_{x,z}$ and $v_{y,z}$ are not in $C_{OUT}(n)$.
Finally, when calculating $C_{NODE}(n)$, the impact values of the vulnerabilities targeting any of the $G_n$ states are taken into account, including the impact values of any possible attacks between the $G_n$ states.

\section*{Acknowledgment}
The authors would like to thank Ambrose Kam and Arlanda Johnson for their invaluable and constructive feedback on the methodology, which helped improve the quality of this study.

\bibliographystyle{unsrt}
\bibliography{references.bib}

\begin{IEEEbiographynophoto}{Joni Herttuainen}
received the B.S. and M.S. degrees in computer science from the Lappeenranta University of technology in 2015, and in 2017, respectively.
He is currently pursuing a D.Sc. degree in Computer Science at Aalto University, Espoo, Finland.

From 2016 to 2019, he was first a Technical Student, and later a Technical Fellow at European Organization for Nuclear Research (CERN).
Between 2020 and 2024 he was a System Specialist in the Blue Brain Project, a brain research initiative, managed by the Swiss Federal Technology Institute of Lausanne (EPFL).
His current research focuses on AI-driven advancements in cybersecurity, including network anomaly detection and computationally enhanced defense mechanisms.
His research also explores the use of AI by adversaries, modelling attack strategies, and conducting quantitative analysis of emerging threats to develop effective defenses.

\end{IEEEbiographynophoto}

\begin{IEEEbiographynophoto}{Vesa Kuikka}
received the Doctor of Military Sciences degree from the National Defence University, Department of Military Technology, in 2021, and the Doctor of Science (Technology) degree from Aalto University, Department of Computer Science, in 2022.

His research has focused on modelling the impact of technologies and systems on military capabilities, including capability optimization, stochastic combat modelling, probability of victory, and network-centric warfare metrics.
This work was published in journals such as Journal of Battlefield Technology, Journal of Military Studies, Journal of Applied Operational Research, Military Operations Research, and Journal of Information Warfare.
His current research focuses on complex network modelling, cyber-related graph analysis, influence and information spreading, community detection, and network resilience.
He has published in journals including Scientific Reports, Physica A, Computational Social Networks, Sensors, Systems, Computation, and Frontiers in Complex Systems, and has contributed to several IEEE conferences.
His work integrates statistical physics, probabilistic modelling, actuarial methods, network science, and systems engineering to analyse cyber, socio-technical, and defense systems.

Dr. Kuikka is a Fellow of the Actuarial Society of Finland (FASF).

\end{IEEEbiographynophoto}

\begin{IEEEbiography}[{\includegraphics[width=1in,height=1.25in,clip,keepaspectratio]{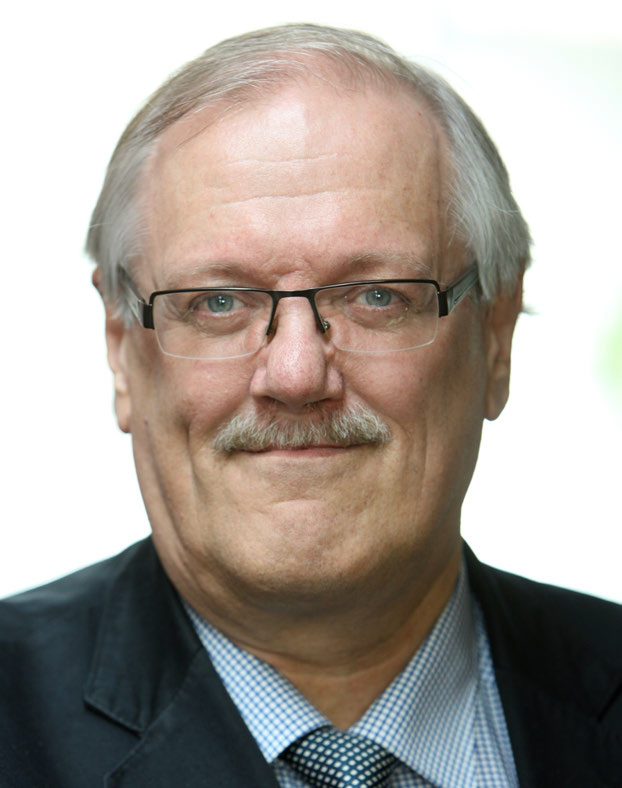}}]{Kimmo K. Kaski}
received the M.Sc. (Tech.) and Lic.Tech. degrees in electrical engineering from the Helsinki University of Technology, Finland, in 1973 and 1977, respectively, and the D.Phil. degree in theoretical physics from the University of Oxford, U.K., in 1981.

He is currently a Professor of computational science with the School of Science, Aalto University, Finland; a Supernumerary Fellow with the Wolfson College, University of Oxford, U.K.; an External Faculty with the Complexity Science Hub, Vienna, Austria; and a Visiting Fellow with The Alan Turing Institute, London, U.K.
His research interests include computational science, statistical physics, complex system science, network science, data science, artificial intelligence, and their applications in various issues of social networks, digital health, wellbeing and resilience.

He is a fellow of the American Physical Society, a fellow and a Chartered Physicist of the Institute of Physics, U.K., a fellow of the Academia Europaea and Finnish Academy of Science and Letters, a fellow of the Finnish Academy of Technical Sciences, and a fellow of the Mexican Academy of Sciences.
\end{IEEEbiography}

\EOD

\end{document}